\documentclass[11pt]{article}
\usepackage{jheppub}
\usepackage{amsmath,amssymb,amsfonts,amsthm,mathtools}
\usepackage{revsymb}
\usepackage{bm}
\usepackage{subfigure}
\usepackage{color}
\usepackage{lineno}

\makeatletter
\def\@fpheader{\relax}
\makeatother

\newcommand{\bra}[1]{\langle #1 |}
\newcommand{\ket}[1]{| #1 \rangle}

\newcommand\be{\begin{equation}}
\newcommand\ee{\end{equation}}
\newcommand\bea{\begin{eqnarray}}
\newcommand\eea{\end{eqnarray}}
\newcommand\ba{\begin{array}}
\newcommand\ea{\end{array}}

\newcommand\bc{\begin{center}}
\newcommand\ec{\end{center}}

\newcommand\comment[1]{}

\newcommand{\X}{\mathbb{X}}
\newcommand{\hX}{\hat{\mathbb{X}}}

\newcommand{\hx}{\mathord{\hat x}}
\newcommand{\tx}{\mathord{\tilde x}}
\newcommand{\htx}{\mathord{\hat{\tilde x}}}

\title{Dynamical Dark Energy, Dual Spacetime, and DESI}

\author[a]{Sunhaeng Hur,}
\author[b]{Vishnu Jejjala,}
\author[c]{Michael J.\ Kavic,}
\author[a]{Djordje Minic,}
\author[a]{Tatsu Takeuchi\,}

\affiliation[a]{Department of Physics, Virginia Tech, Blacksburg, VA 24061, U.S.A.}
\affiliation[b]{Mandelstam Institute for Theoretical Physics, School of Physics and NITheCS\\
University of the Witwatersrand, Johannesburg, WITS 2050, South Africa}
\affiliation[c]{Department of Chemistry and Physics, SUNY Old Westbury, Old Westbury, NY 11568, U.S.A.}

\emailAdd{sunhh@vt.edu}
\emailAdd{v.jejjala@wits.ac.za}
\emailAdd{kavicm@oldwestbury.edu}
\emailAdd{dminic@vt.edu}
\emailAdd{takeuchi@vt.edu}

\abstract{
In this paper we discuss possible consequences of a manifestly non-commutative and $T$-duality covariant formulation of string theory on dark energy, when the correspondence between short distance (UV) and long distance (IR) physics is taken into account. We demonstrate that the dark energy is dynamical, \textit{i.e.}, time-dependent, and we compute the allowed values of $w_0$ and $w_a$, given by $w(a) = w_0+w_a(1-a)$, which compare favorably to the most recent observations by DESI. From this point of view, the latest results from DESI might point to a fundamentally new understanding of quantum spacetime in the context of quantum gravity.
}

\keywords{dynamical dark energy, equation-of-state parameter, Hubble parameter, infinite statistics\\

\noindent \textsc{Comments:} This extends and updates \href{https://arxiv.org/abs/2202.05266}{arXiv:2202.05266}.}

\begin{document}

\maketitle
\parskip=.5\baselineskip

\section{Introduction}\label{sec:intro}

\noindent
Driven by an unknown dark energy, the expansion of the Universe is accelerating.
To date, the simplest explanation that accommodates the data is adding a cosmological constant term to the Einstein--Hilbert action.
This imposes a constant vacuum energy density that in the current epoch is the leading contribution to the Friedmann equation and yields, at late times, an exponential expansion.
Alternatively, mimicking the inflationary phase in the early history of the Universe, quintessence fields in a potential (for a review and references, consult, for example~\cite{Tsujikawa:2013fta})  may produce de Sitter-like behavior.
In the latter scenario, the dark energy is explicitly dynamical.

From a purely observational perspective, the cosmological constant scenario fits the data well and predicts an equation-of-state parameter $w\approx -1$ that is consistent with measurements of type Ia supernov\ae, the cosmic microwave background (CMB), and large scale structure~\cite{Peebles:2024txt, Perivolaropoulos:2021jda}.
Despite passing these consistency checks, the cosmological constant hypothesis suffers from famous fine tuning and coincidence problems, and de Sitter spacetimes --- even metastable ones --- are notoriously difficult (but perhaps, not impossible) to realize in quantum gravity that also includes the observed visible matter sector and allows for a dark matter sector~\cite{Freidel:2022ryr, Berglund:2022qsb, Berglund:2023gur}.
(See also~\cite{Danielsson:2018ztv, Obied:2018sgi}.)
The well publicized tension between high redshift measurements (around $67$ km/s/Mpc) and low redshift measurements (around $73$ km/s/Mpc) of the Hubble parameter $H_0$ can be accounted for by a dynamical dark energy~\cite{DiValentino:2021izs}.
The difficulty with quintessence models is that specifying their phenomenology introduces a large number of new parameters into the effective action that are themselves finely tuned in order to ensure that the transition to the de Sitter phase occurred quite recently in the history of the Universe~\cite{DiValentino:2021izs}.
Moreover, significant deviations from $w=-1$ are unsupported by data.
Current and near-future surveys such as DESI~\cite{DESI:2025zgx}, Euclid\footnote{https://www.esa.int/Science\_Exploration/Space\_Science/Euclid\_overview} 
\cite{EuclidTheoryWorkingGroup:2012gxx}, and the Vera Rubin Observatory\footnote{https://www.lsst.org/science/dark-energy} search for signals of dark energy and herald an era of precision cosmology that can distinguish between these possibilities.
In this work we concentrate on the recent exciting results announced by DESI~\cite{DESI:2025zgx}.

In particular, we present an approach to dynamical dark energy based on general theoretical considerations.
We make use of a recent development in understanding string theory in an intrinsically non-commutative T-duality covariant formulation~\cite{Freidel:2013zga, Freidel:2015uug, Freidel:2016pls, Freidel:2017xsi, Freidel:2017wst}.
In this formulation, apart from familiar spacetime $x$, there exists a notion of dual spacetime $\tilde{x}$, which does not commute with $x$, $[x, \tilde{x}]= i \lambda^2$, with $\lambda$ a fundamental length parameter, making the very notion of spacetime polarization observer dependent.
There are important implications of this new perspective, one of which is that the curvature of dual spacetime, to leading order in $\lambda$, represents the cosmological constant in Einstein's gravitational equations in the spacetime $x$.
We use this observation, as well as the statistics of infinite matrices $x$ and $\tilde{x}$ (the so called infinite, or quantum distinguishable, statistics) that satisfy the Heisenberg algebra $[x, \tilde{x}]= i \lambda^2$, in order to formulate a generic model of a dynamical dark energy, compatible in a suitable limit with Einstein's cosmological constant.
From this point of view, dark energy originates from the curvature of the dual spacetime, which upon quantization, is made out of ``atoms,'' that obey quantum distinguishable, or quantum Boltzmann, or infinite, statistics.
In particular, by using this viewpoint on dark energy and by utilizing generic relations between ultraviolet (UV) and infrared (IR) cutoffs, we compute the equation-of-state parameter for dynamical dark energy and favorably compare it with recent measurements from DESI~\cite{DESI:2025zgx}.

The paper is organized as follows.
In Section~\ref{sec:dualspacetime}, we first give a basic mathematical motivation for our approach.
In Section~\ref{sec:w}, we then discuss dynamical dark energy and the calculation of the equation-of-state parameter and compare it to the most recent observations made by DESI.
We conclude in Section~\ref{sec:discussion} with a brief discussion of our results.
The theoretical background to this paper is collected in the Appendices.

\section{Motivation: Dual spacetime models}\label{sec:dualspacetime}

In this section, we use the manifestly T-dual formulation of string theory to motivate the expression
\begin{equation}
\rho_{\Lambda}(a) 
\;=\; \dfrac{3H_0^2}{8\pi G_N}\,\Omega_\Lambda(a)
\;=\; \dfrac{\Lambda(a)}{8\pi G_N} \;=\;
\int_{\tilde{E}_\mathrm{IR}(a)}^{\tilde{E}_\mathrm{UV}(a)}d\tilde{E}\;I(\tilde{E})\;,
\end{equation}
for the dynamical dark energy density, where $a$ is the scale factor of Friedmann--Robertson--Walker (FRW) cosmology (in our Universe), and $I(\tilde{E})$ is the density of states (in the dual spacetime).
Note that what follows is only a motivation, and not a suggestion for a detailed model that underlies the discussion and the results of this paper.

The starting point for the motivation of our calculation is the manifestly T-dual action proposed by Tseytlin in~\cite{Tseytlin:1990hn}.
This action can be found in a limit of an intrinsically non-commutative and T-duality covariant formulation of string theory called the \textit{metastring}~\cite{Freidel:2013zga, Freidel:2015uug, Freidel:2016pls, Freidel:2017xsi, Freidel:2017wst}, which provides a new perspective on quantum gravity as a gravitized quantum theory~\cite{Hubsch:2024agh}, with many other observational implications~\cite{Hubsch:2024agh, Berglund:2023vrm}.
The explicit action involves two spacetime covariant labels $x$ and $\tilde{x}$, which do not commute in the metastring formulation $[x, \tilde{x}] = i \lambda^2$, where $\lambda$ is a effective length.
However, to first order in $\lambda$, the two labels $x$ and $\tilde{x}$ commute and the Tseyltin action reads as follows:
\bea
S & = & - \int_{x}\int_{\tx} \sqrt{-g(x)} \sqrt{-\tilde{g}(\tx)} 
\,\Bigl[\, 
  R(x) + \tilde{R}(\tx) + \cdots
\,\Bigr]
\cr
& = & -\int_{x}\sqrt{-g(x)}\left[
R(x)\int_{\tx}\sqrt{-\tilde{g}(\tx)} 
+\int_{\tx}\sqrt{-\tilde{g}(\tx)}\;\tilde{R}(\tx)
+\cdots
\right]
\;.
\label{eq:TsSd}
\eea
The $\tx$-integration of the first term of~\eqref{eq:TsSd} defines the gravitational constant $G_N$
in the $x$-spacetime
\be
1/G_N \;\sim\; \int_{\tx} \sqrt{-\tilde{g}(\tx)} \;,
\label{OneOverGN}
\ee
and that of the second term produces a positive cosmological constant $\Lambda >0$ (dark energy)
\be
\Lambda/G_N \;\sim\; \int_{\tx} \sqrt{-\tilde{g}(\tx)}\,\tilde{R}(\tx) \;.
\label{eq:LambdaOverGN}
\ee
Thus, the weakness of gravity is determined by the size of the canonically conjugate dual $\tx$-spacetime,  while the smallness of the cosmological constant is given by its curvature $\tilde{R}$. 
Note that this expression also implies that $\Lambda/G_N$ will evolve together with the evolution of the $\tx$-spacetime.
The expression for $G_N$ has to be regularized, in order to be properly defined.
Thus, in principle, $G_N$ would depend on the regulator, and also, it would change (flow) with the change of the regulator.
The physical $G_N$ would in principle be identifiable by the fixed point value of that flow.
Similarly, because the covariant spacetime labels $x$ and $\tilde{x}$ do not commute, they are not independent and thus, the change of the curvature in $\tilde{x}$, which is the origin of $\Lambda$, would, in principle, imply that $\Lambda$ does change with the volume of $x$, of alternatively, with the scale factor $a$ of the $x$ spacetime, as well.
This is the motivation for looking at how $\Lambda(a)$ changes with the scale factor $a$ of the observed spacetime.
Once again, this change (flow), might have a fixed point, which would be identified with the vacuum energy density that corresponds to the cosmological constant in Einstein's equations.
We note in passing that all flows in the non-commutative context are expected to have UV/IR mixing and a double renormalization group that ensures the existence of the continuum limit and a generically self-dual fixed point (with respect to the UV and IR cutoffs)~\cite{Grosse:2004yu}.
This logic is in accord with the general intuition that cosmological evolution can be modeled as an effective renormalization group evolution of some appropriate dual description of gravitational dynamics in asymptotically de Sitter backgrounds, as discussed, for example, in~\cite{Strominger:2001gp, Balasubramanian:2001nb}.

The vacuum energy density is
\begin{equation}
\rho_{\Lambda}(a) \;=\; \dfrac{\Lambda(a)}{8\pi G_N} \;,
\end{equation}
which in our consideration is proportional to the integral given in~\eqref{eq:LambdaOverGN}.
Given that the vacuum energy density can be considered as the sum of the vacuum fluctuation energies of all the quantum states,
we can write
\begin{equation}
\rho_{\Lambda}(a) \;=\; \int_{\tilde{E}_\mathrm{IR}(a)}^{\tilde{E}_\mathrm{UV}(a)}d\tilde{E}\,I(\tilde{E})\;,
\end{equation}
where $\tilde{E}$ is the energy of the quantum fluctuations in the dual $\tilde{x}$-spacetime, $I(\tilde{E})$ is the density of states in $\tilde{E}$-space, which depends on the curvature $\tilde{R}$ of the $\tilde{x}$-spacetime in some fashion,
and $\tilde{E}_\mathrm{IR}(a)$ and $\tilde{E}_\mathrm{UV}(a)$ are the IR and UV cutoffs which we assume to be $a$-dependent in general.
In principle, there is a contribution from visible and dark matter fields here, and that is included in the observed value of the vacuum energy that contributes to the cosmological constant.
The volume of spacetime $x$ depends on $a$, and if we convert (via the ``uncertainty" relation implied by the fundamental non-commutativity of $x$ and $\tilde{x}$, $\delta x \delta \tilde{x} \sim \lambda^2$) the volume of spacetime $x$ to the volume of spacetime $\tilde{x}$, and this in turn, via the usual Heisenberg uncertainty relation to the energy in the dual spacetime $\tilde{x}$, we obtain the above equation, in which the integration is performed over the dual energy $\tilde{E}$.  These ideas are elaborated in Appendix~\ref{sec:qsqg}.

\section{Dynamical dark energy and the equation-of-state parameter}\label{sec:w}

\subsection{General properties}

In our model, dark energy stems from the curvature of the dual spacetime (a generic feature of a general formulation of string theory), which upon quantization, is made out of ``atoms,'' or ``quanta,'' that obey quantum distinguishable, or infinite, statistics.
In this central section of the paper, we use this view of dark energy to construct a rather general model of, in principle, time-dependent dark energy, based on the notion of infinite statistics and the UV/IR relation (another generic property of string theory).
We then show that our theoretical considerations favorably compare to the most recent observations of DESI~\cite{DESI:2025zgx}, which provide constraints on dynamical dark energy in the Chevallier--Polarski--Linder (CPL) parametrization~\cite{Chevallier:2000qy,Linder:2002et,dePutter:2008wt}:
\begin{equation}
w(a) \;=\; w_0 + w_a(1-a)\;.
\label{CPL}
\end{equation}

Without making any assumptions about the $a$-dependence of the UV and IR cutoffs in $\tilde{E}$-space, we can say a few things about the dark energy equation-of-state parameter $w(a)$, using the
derivation provided in Appendix~\ref{app:w}.
Starting from 
\begin{equation}
\rho_{\Lambda}(a) 
\;=\; \dfrac{3H_0^2}{8\pi G_N}\,\Omega_\Lambda(a)
\;=\; \dfrac{\Lambda(a)}{8\pi G_N} \;=\;
\int_{\tilde{E}_\mathrm{IR}(a)}^{\tilde{E}_\mathrm{UV}(a)}d\tilde{E}\,I(\tilde{E})\;,
\end{equation}
we find, using~\eqref{wafromOmegaLambda},
\begin{eqnarray}
w(a) & = & -1 -\dfrac{a}{3}\dfrac{d}{da}\log\Omega_\Lambda(a) 
\vphantom{\bigg|}\cr
& = & -1 -\dfrac{a}{3\Lambda(a)}\dfrac{d\Lambda(a)}{da} 
\vphantom{\bigg|}\cr
& = & -1 -\dfrac{8\pi G_N a}{3\Lambda(a)}\dfrac{d}{da}
\bigg[\int_{\tilde{E}_\mathrm{IR}(a)}^{\tilde{E}_\mathrm{UV}(a)}d\tilde{E}\;I(\tilde{E})\bigg]
\vphantom{\Bigg|}\cr
& = & -1 -\frac{8\pi G_{N}a}{3\Lambda(a)}
\bigg[
 I(\tilde{E}_\mathrm{UV}(a))\frac{d\tilde{E}_\mathrm{UV}(a)}{da}
-I(\tilde{E}_\mathrm{IR}(a))\frac{d\tilde{E}_\mathrm{IR}(a)}{da}
\bigg]
\;.
\vphantom{\Bigg|}
\label{waGeneric}
\end{eqnarray}
Therefore, $w(a)<-1$ if
\begin{equation}
I(\tilde{E}_\mathrm{UV}(a))\frac{d\tilde{E}_\mathrm{UV}(a)}{da}
\;>\;
I(\tilde{E}_\mathrm{IR}(a))\frac{d\tilde{E}_\mathrm{IR}(a)}{da}
\;.
\end{equation}
and $w(a)>-1$ if 
\begin{equation}
I(\tilde{E}_\mathrm{UV}(a))\frac{d\tilde{E}_\mathrm{UV}(a)}{da}
\;<\;
I(\tilde{E}_\mathrm{IR}(a))\frac{d\tilde{E}_\mathrm{IR}(a)}{da}
\;.
\end{equation}

\subsection{UV--IR correspondence}

We now make some assumptions about the $a$-dependence of $\tilde{E}_\mathrm{UV}(a)$ and $\tilde{E}_\mathrm{IR}(a)$.
Given the general structure of the metastring formulation~\cite{Freidel:2013zga, Freidel:2015uug, Freidel:2016pls, Freidel:2017xsi, Freidel:2017wst} and its metaparticle limit~\cite{Freidel:2018apz, Freidel:2021wpl}, as a first, naive, attempt, we simply assume that
\begin{equation}
E_\mathrm{IR}\tilde{E}_\mathrm{UV} \;=\; E_\mathrm{UV}\tilde{E}_\mathrm{IR} \;=\; \mu\;,
\label{UVIR}
\end{equation}
where the symbols without the tilde refer to the cutoffs in $x$-spacetime, and 
$\sqrt\mu$ a reference energy scale related to the mixing of UV and IR degrees of freedom due to non-locality~\cite{Jejjala:2020lhg}.
In anti-de Sitter (AdS) space, the low energy cutoff $E_\mathrm{IR}$ 
is determined by the radius of curvature~\cite{Heemskerk:2009pn}.
In an approximately de Sitter (dS) space, this scale is set by the cosmological horizon~\cite{Berglund:2022qsb}. 
Thus,
\begin{equation}
E_\text{IR}(a) \;=\; \frac{E_0}{a}  
\qquad\to\qquad \tilde{E}_\mathrm{UV}(a) \;=\; \dfrac{\mu}{E_\mathrm{IR}(a)} \;=\; \dfrac{\mu a}{E_0}
\;,
\label{eq:eir}
\end{equation}
where $E_0$ is the infrared cutoff today, at $a=1$.
Note that $\lim_{a\to 0}E_\text{IR}(a)=\infty$ since the size of the $x$-Universe collapses to zero as $a\to 0$,
but this is unphysical so we should think of the expression as effective away from $a=0$.
This assumption leads to
\begin{equation}
\dfrac{d\tilde{E}_\mathrm{UV}(a)}{da} \;=\; \dfrac{\mu}{E_0} \;>\; 0\;.
\end{equation}
If $G_N$ is constant, then $E_\mathrm{UV} = E_\mathrm{Planck}$ in the $x$-spacetime, which implies that $\tilde{E}_\mathrm{IR}(a)=\text{constant}$ in the $\tilde{x}$-spacetime, which would imply $w(a)<-1$ without making further assumptions.
But if~\eqref{OneOverGN} implies an $a$-dependent Planck length it could lead to an $a$-dependent $\tilde{E}_\mathrm{IR}(a)$.
However, even if $\tilde{E}_\mathrm{IR}$ were not constant, as long as it is decreasing with $a$, then $w(a)<-1$ without making further assumptions. (Note that the double renormalization group flows of an intrinsically non-commutative set-up in principle allow to have $\tilde{E}_\mathrm{UV}(a)$ decrease with $a$, and have $\tilde{E}_\mathrm{IR}(a)$ increase with $a$.
This would allow $w>-1$. Also, as argued in the second section,
the $G_N$ would be, in principle, finite at such a self-dual fixed point.)
We will discuss a more general relation between the UV and IR cutoffs in what follows. But first we want to examine the general consequences of infinite statistics on our analysis.

\subsection{Consequences of infinite statistics}\label{sec:InfStat}

\begin{figure}[t]
\begin{center}
\includegraphics[height=4cm]{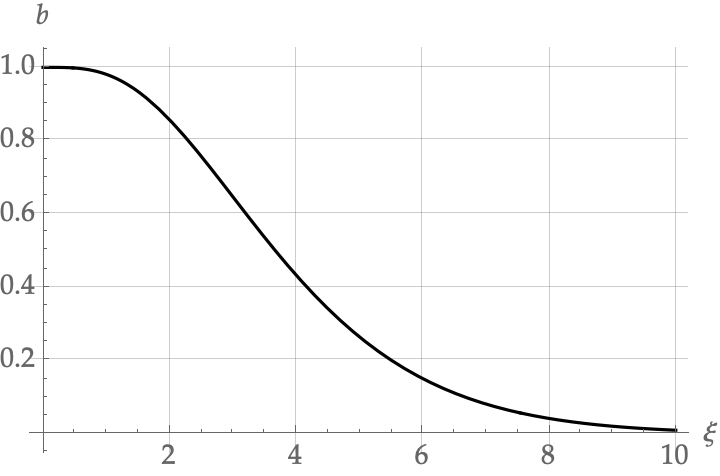}
\caption{The function $b(\xi)$ defined in Eq.~\eqref{eq:params1}.}
\label{bFig}
\end{center}
\end{figure}

Infinite statistics (or quantum distinguishable, or quantum Boltzmann statistics) is the most general statistics consistent with Lorentz invariance and non-locality --- see Appendix~\ref{sec:infinitestatistics}. Infinite statistics is naturally realized in matrix models, and given the fact that a non-perturbative metastring formulation of string theory involves matrices (as reviewed in Appendix~\ref{sec:qsqg}), infinite statistics is the natural statistics of spacetime ``quanta.'' In our model, dark energy comes from the curvature of dual spacetime (another generic feature of the metastring), and thus its ``quanta'' have to obey infinite statistics.
Therefore, in an effective four-dimensional background, infinite statistics  tells us that the dark energy quanta follow (quantum) Boltzmann statistics, which we express in terms of the Wien distribution, in order to have proper normalization of the vacuum energy,
\be
I_\text{DE}(\tilde{E}, E_0) \;=\; A\, \tilde{E}^3\, e^{-B \tilde{E}/E_0} \;, 
\label{eq:ide}
\ee
with $A$ and $B$ being dimensionless constants~\cite{Jejjala:2007hh}.
The observed vacuum energy density arises from integrating the previous expression:
\be
\rho_\Lambda (a) 
\;=\; \frac{\Lambda(a)}{8\pi G_\text{N}} 
\;=\; \int_{\tilde{E}_\mathrm{IR}(a)}^{\tilde{E}_\mathrm{UV}(a)} d\tilde{E}\ I_\text{DE}(\tilde{E}, E_0) 
\;.
\label{eq:rhovac}
\ee
(In this convention, $\Lambda$ has units of $[\text{time}]^{-2}$.)
Performing the integral, we find
\begin{equation}
\rho_\Lambda(a) 
\;=\; \frac{\Lambda(a)}{8\pi G_\text{N}} 
\;=\; \rho_* \Big[b\big(\xi_\text{IR}(a)\big)-b\big(\xi_\text{UV}(a)\big)\Big] \;, 
\label{eq:Lofz}
\end{equation}
where we have defined 
\begin{equation}
\xi_\text{UV}(a) \;=\; \dfrac{B\tilde{E}_\text{UV}(a)}{E_0}\;,\qquad
\xi_\text{IR}(a) \;=\; \dfrac{B\tilde{E}_\text{IR}(a)}{E_0}\;,
\end{equation}
which are dimensionless, and
\begin{equation}
\rho_* \;=\; \frac{6A}{B^4}\, E_0^4 \;, \qquad 
b(\xi) \;=\; \left( 1 + \xi + \frac{1}{2} \xi^2 + \frac{1}{6} \xi^3 \right) e^{-\xi} \;.
\label{eq:params1}
\end{equation}
The scaling $\rho_* \sim E_0^4$ seen in~\eqref{eq:params1} is analogous to the Stefan--Boltzmann law, which tells us that the power of the radiation emitted by a blackbody scales as $T^4$.
The function $b(\xi)$ is shown in Figure~\ref{bFig}.
From~\eqref{eq:Lofz}, we can see that
if $\xi_\text{UV}(0)=\xi_\text{IR}(0)$, then $\Lambda(0)=0$ since the integration range vanishes in that limit.

From~\eqref{waGeneric}, the expression for $w(a)$ becomes
\begin{eqnarray}
w(a) & = & 
-1 -\frac{a}{3\rho_\Lambda(a)}
\bigg[
 I(\tilde{E}_\mathrm{UV}(a))\frac{d\tilde{E}_\mathrm{UV}(a)}{da}
-I(\tilde{E}_\mathrm{IR}(a))\frac{d\tilde{E}_\mathrm{IR}(a)}{da}
\bigg]
\vphantom{\Bigg|}\cr
& = &
-1 -\dfrac{1}{18\big[b_\text{IR}(a)-b_\text{UV}(a)\big]}
\bigg[
 \xi_\text{UV}^3\, e^{-\xi_\text{UV}}\bigg( a\,\dfrac{d\xi_\text{UV}}{da} \bigg)
-\xi_\text{IR}^3\, e^{-\xi_\text{IR}}\bigg( a\,\dfrac{d\xi_\text{IR}}{da} \bigg)
\bigg]\;,
\cr & &
\label{wageneric}
\end{eqnarray}
where $b_\text{UV}(a) = b(\xi_\text{UV}(a))$ and $b_\text{IR}(a) = b(\xi_\text{IR}(a))$.
Beyond this point we need the $a$-dependences of $\xi_\text{UV}$ and $\xi_\text{IR}$.

\bigskip

\begin{figure}[ht]
\begin{center}
\includegraphics[height=4cm]{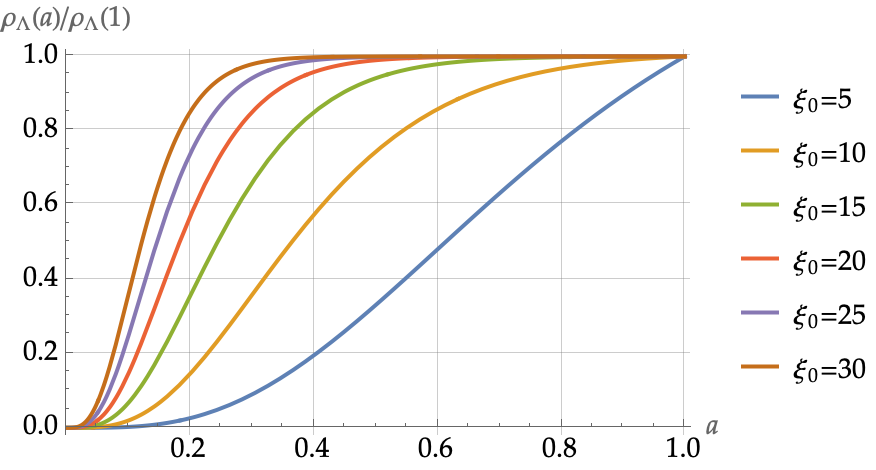}
\caption{The dark energy density $\rho_\Lambda(a)/\rho_\Lambda(1)$ as a function of $a$ for various values of $\xi_0$.}
\label{rhofig1}
\end{center}
\end{figure}

\begin{figure}[ht]
\begin{center}
\includegraphics[height=4cm]{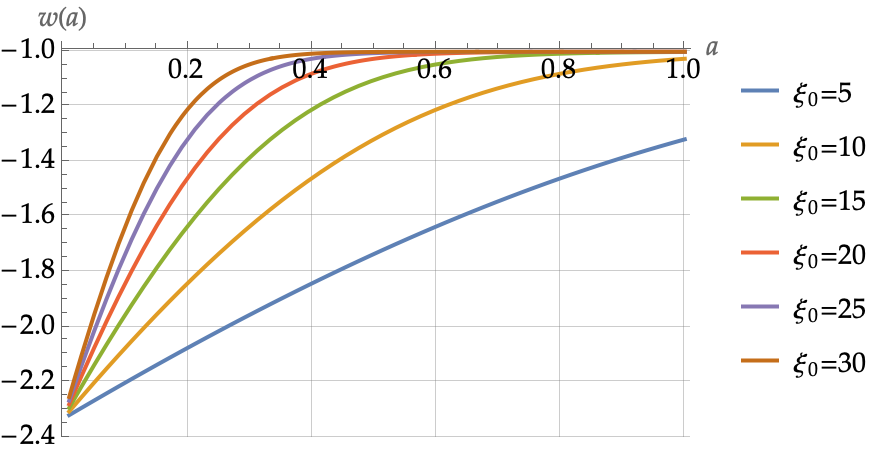}
\caption{The equation-of-state parameter $w(a)$ as a function of $a$ for various values of $\xi_0$.}
\label{wafig1}
\end{center}
\end{figure}

\begin{figure}[ht]
\begin{center}
\includegraphics[height=4cm]{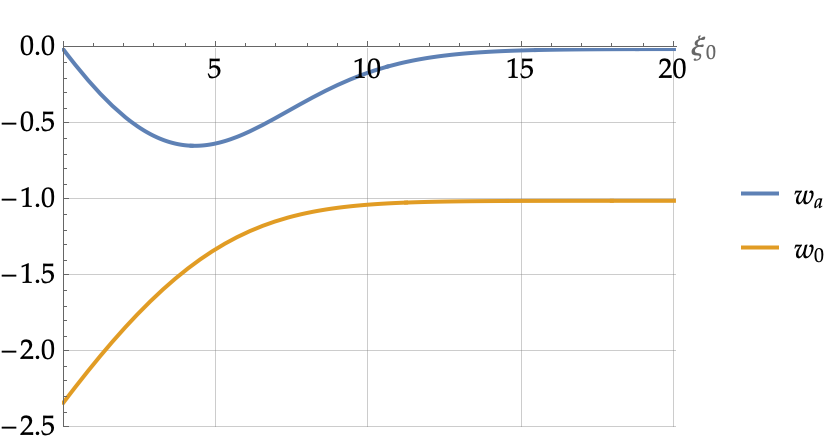}
\caption{$w_0=w(1)$ and $w_a=-w'(1)$ as functions of $\xi_0$.}
\label{w0waFig}
\end{center}
\end{figure}

Motivated by~\eqref{eq:eir}, let us try
\begin{equation}
\tilde{E}_\mathrm{UV}(a) \;=\; \dfrac{\mu}{E_0}\,a\;,\qquad
\tilde{E}_\mathrm{IR}(a) \;=\; 0\;.
\end{equation}
Then
\begin{equation}
\xi_\text{UV}(a) \;=\; \dfrac{B\mu}{E_0^2}\,a \;\equiv\; \xi_0\,a\;,\qquad
\xi_\text{IR}(a) \;=\; 0\,,
\end{equation}
and
\begin{equation}
a\,\dfrac{d\xi_\text{UV}(a)}{da} \;=\; \xi_0\,a \;=\; \xi_\text{UV}(a)\;,\qquad
a\,\dfrac{d\xi_\text{IR}(a)}{da} \;=\; 0\;.
\end{equation}
We obtain
\begin{equation}
\rho_\Lambda(a) 
\;=\; \frac{\Lambda(a)}{8\pi G_\text{N}} 
\;=\; \rho_* \Big[1-b\big(\xi_0 a\big)\Big] \;, 
\end{equation}
and
\begin{equation}
w(a) \;=\; -1 - \frac{\xi^4 e^{-\xi}}{18\{1-b(\xi)\}} \;, \qquad \xi \;=\; \xi_0\, a \;.
\end{equation}
Clearly, $w(a)<-1$ for all $a$ as anticipated. We show the behavior of $\rho_\lambda(a)$ and $w(a)$ for several choices of $\xi_0$ in Figures~\ref{rhofig1} and~\ref{wafig1}.
The derivative of $w(a)$ is
\begin{equation}
\dfrac{dw(a)}{da}
\;=\; \dfrac{d\xi}{da}\dfrac{dw(\xi)}{d\xi}
\;=\; -\dfrac{\xi_0}{18}\bigg[\dfrac{(4-\xi)\xi^3 e^{-\xi}}{\{1-b(\xi)\}} - \dfrac{\xi^7 e^{-2\xi}}{6\{1-b(\xi)\}^2}
\bigg]
\;.
\end{equation}
Therefore, adopting the CPL parameterization, Eq.~\eqref{CPL},
we find 
\begin{eqnarray}
w_0 \;=\; w(1) & = & -1 - \frac{\xi_0^4 e^{-\xi_0}}{18\{1-b(\xi_0)\}}\;,
\vphantom{\Bigg|}\cr
w_a \;=\; -w'(1) & = & \dfrac{\xi_0^4 e^{-\xi_0}}{18\{1-b(\xi_0)\}}\bigg[(4-\xi_0) - \dfrac{\xi_0^4 e^{-\xi_0}}{6\{1-b(\xi_0)\}}\bigg] 
\cr
& = & -(4-\xi_0)(w_0+1)-3(w_0+1)^2
\;.\vphantom{\Bigg|}
\label{w0wa}
\end{eqnarray}
The $\xi_0$ dependence of $w_0$ and $w_a$ are shown in Figure~\ref{w0waFig}. 

\begin{figure}[ht]
\begin{center}
\includegraphics[height=4cm]{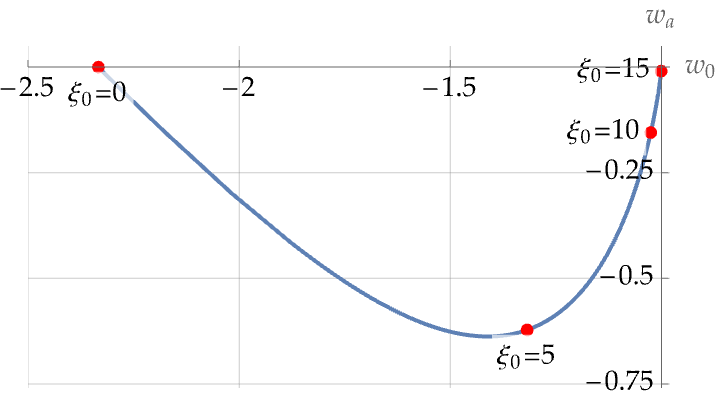}
\caption{Trajectory of the point $(w_0,w_a)$ as $\xi_0$ is changed.}
\label{w0waFig2}
\end{center}
\end{figure}

\noindent
Note that $w_0\le -1$ and $w_a\le 0$ for all $\xi_0$.
The graphs of $w(a)$ for different values of $\xi_0$ shown previously are just the graph of $w_0$ between $0$ and $\xi_0$
rescaled horizontally to fit between 0 and 1.
In the limit $\xi_0\to\infty$, we find $(w_0,w_a)\to(-1,0)$ which recovers the cosmological constant.
For $\xi_0\ll 1$, we find
\begin{eqnarray}
w_0 & = & -\dfrac{7}{3} + \dfrac{4}{15}\xi_0 - \dfrac{2}{225}\xi_0^2 - \dfrac{2}{2625}\xi_0^3 + O(\xi_0^4)\;, 
\vphantom{\bigg|}\cr
w_a & = & -\dfrac{4}{15}\xi_0 + \dfrac{4}{225}\xi_0^2 + \dfrac{2}{875}\xi_0^3 + O(\xi_0^4)\;,
\vphantom{\bigg|}
\end{eqnarray}
so $(w_0,w_a)\to(-7/3,0)$ in the $\xi_0\to 0$ limit. 
As $\xi_0$ is increased from 0 to $\infty$, 
$(w_0,w_a)$ follows the trajectory shown in Figure~\ref{w0waFig2}.
Compare with Figure~11 of Ref.~\cite{DESI:2025zgx} from DESI.
The DESI analysis of~\cite{DESI:2025zgx} uses priors which allow $w_0$ to be less than $-1$, 
so this model is clearly disfavored.

\noindent

\bigskip
Let us also look at a case in which both $\tilde{E}_\text{UV}$ and $\tilde{E}_\text{IR}$ are $a$-dependent.
Motivated by~\eqref{UVIR}, let us try
\begin{equation}
\tilde{E}_\text{UV} \;=\; \dfrac{\mu}{E_0}\big[\alpha + (1-\alpha)a\big]\;,\qquad
\tilde{E}_\text{IR} \;=\; \dfrac{\mu}{E_\text{Planck}}\dfrac{1}{[\beta+(1-\beta)a\big]}\;,
\label{alphabeta}
\end{equation}
where 
\begin{equation}
\alpha\beta \;=\; \dfrac{E_0}{E_\text{Planck}}
\end{equation}
so that 
\begin{equation}
\tilde{E}_\text{UV}(0) \;=\; \tilde{E}_\text{IR}(0)
\;,
\end{equation}
and $\tilde{E}_\text{UV}(a)$ increases with $a$, while $\tilde{E}_\text{IR}(a)$ decreases with $a$.
Furthermore, let us assume
\[
\alpha \,=\, \beta \,=\, \sqrt{\dfrac{E_0}{E_\text{Planck}}} \,\equiv\, k\,,\qquad
\tilde{E}_\text{UV}(0) \,=\, \tilde{E}_\text{IR}(0)
\,=\, \dfrac{\mu}{\sqrt{E_0 E_\text{Planck}}}\;, 
\]
to reduce the number of free parameters. Then
\begin{eqnarray}
\xi_\text{UV}(a) & = & \frac{B\mu}{E_{0}^{2}}[k+(1-k)a] \;=\; \xi_0[k+(1-k)a] \;,
\vphantom{\Bigg|}\cr
\xi_\text{IR}(a) & = & \frac{B\mu}{E_{0}E_\text{Planck}}[k+(1-k)a]^{-1}
\cr
& = & \xi_0\frac{E_{0}}{E_\text{Planck}}[k+(1-k)a]^{-1}
\;=\; \xi_0\,k^{2}[k+(1-k)a]^{-1} \;,
\vphantom{\Bigg|}\cr
& &
\label{alphabetak}
\end{eqnarray}
where $\xi_0 = B\mu/E_0^2$ as before. 
Looking at these expressions, it is clear that $\xi_\text{UV}(a)\approx \xi_0 a$ $\xi_\text{IR}(a)\approx 0$ when 
$\alpha=\beta=k$ is small,
so there is essentially no difference from what we had before. 
Relaxing the condition $\alpha=\beta$ would not change things much either since both parameters will be tiny.

\subsection{Alternative UV--IR correspondence }

\begin{figure}[ht]
\begin{center}
\includegraphics[height=4cm]{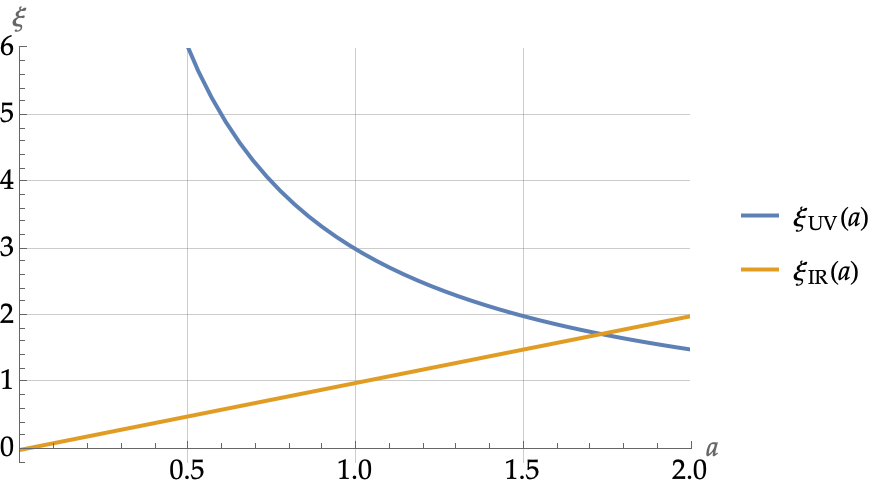}
\caption{The behaviors of $\xi_\text{IR}(a)$ and $\xi_\text{UV}(a)$ for the case $\xi_\text{IR}^{(1)}=1$, $\xi_\text{UV}^{(1)}=3$.
The cutoffs meet at $a=\sqrt{\xi_\text{UV}^{(1)}/\xi_\text{IR}^{(1)}}=\sqrt{3}$.}
\label{cutoffplot}
\end{center}
\end{figure}

\noindent
In this section we consider a more general relation between the UV and IR cutoffs. The generic formulation of metastring theory~\cite{Freidel:2013zga, Freidel:2015uug, Freidel:2016pls, Freidel:2017xsi, Freidel:2017wst} as well as its metaparticle limit~\cite{Freidel:2018apz, Freidel:2021wpl},
indeed allows for a more general 
relation between UV and IR cutoffs as implied by a bi-orthogonal metric that implements T-duality in the metastring formulation~\cite{Freidel:2013zga, Freidel:2015uug, Freidel:2016pls, Freidel:2017xsi, Freidel:2017wst}, and in its metaparticle limit~\cite{Freidel:2018apz, Freidel:2021wpl}.
In particular, the relation between the momenta $p$ and dual momenta $\tilde{p}$, and thus, between the UV and IR cutoffs in the observed spacetime and its dual~\cite{Freidel:2018apz, Freidel:2021wpl} can be written more generally as
\begin{equation}
E_\text{IR}\tilde{E}_\text{IR}\;=\;\mu_\text{IR}\;,\qquad 
E_\text{UV}\tilde{E}_\text{UV}\;=\;\mu_\text{UV}\;,
\label{UVIR3}
\end{equation}
where $\mu_\text{IR}< \mu_\text{UV}$. Note that this makes perfect sense, given the general relation between the momenta $p$ and their duals $\tilde{p}$ in the metaparticle limit,
$p \tilde{p} =\mu$, where the value of $\mu$ is in general different for IR and UV, and where the IR and UV limits are consistently implemented in this relation, as opposed to our first, naive guess that implied $w < -1$.
Also, we have
\begin{equation}
E_\text{IR}(a)\;=\;\frac{E_{0}}{a}\;,\quad E_\text{UV}(a)\;=\;E_\text{Planck}\,a\;.
\label{EIREUV}
\end{equation}
Then
\begin{equation}
\tilde{E}_\text{IR}(a)\;=\;\dfrac{\mu_\text{IR} a}{E_{0}}\;,\qquad 
\tilde{E}_\text{UV}(a)\;=\;\dfrac{\mu_\text{UV}}{E_\text{Planck}a}\;,
\label{UVIRcutoffs3}
\end{equation}
that is, $\tilde{E}_\text{IR}(a)$ increases with $a$ while
$\tilde{E}_\text{UV}(a)$ decreases with $a$.
We want 
\begin{equation}
\tilde{E}_\text{IR}(0)\;<\;\tilde{E}_\text{UV}(0)\;,
\end{equation}
and 
$\tilde{E}_\text{IR}(a)$ and $\tilde{E}_\text{UV}(a)$ to meet at $a>1$:
\begin{eqnarray}
\tilde{E}_\text{IR}(a) & = & \tilde{E}_\text{UV}(a) \cr
& \downarrow & \cr
\dfrac{\mu_\text{IR} a}{E_{0}}
& = & \dfrac{\mu_\text{UV}}{E_\text{Planck}a} \cr
& \downarrow & \cr
1 \;<\; a^2 & = & \dfrac{\mu_\text{UV}}{\mu_\text{IR}}\dfrac{E_0}{E_\text{Planck}} 
\;,
\end{eqnarray}
so we must have
\begin{equation}
\dfrac{\mu_\text{UV}}{\mu_\text{IR}} \;>\; \dfrac{E_\text{Planck}}{E_0} \;\gg\; 1\;.
\end{equation}
That is, the ratio $\mu_\text{UV}/\mu_\text{IR}$ must be huge.
We define
\begin{eqnarray}
\xi_\text{IR}(a) & = & \dfrac{B\tilde{E}_\text{IR}(a)}{E_{0}}\;=\;\frac{B\mu_\text{IR}}{E_{0}^{2}}\,a \;\equiv\; \xi_\text{IR}^{(1)}a\;,\cr  
\xi_\text{UV}(a) & = & \dfrac{B\tilde{E}_\text{UV}(a)}{E_{0}}\;=\;\frac{B\mu_\text{UV}}{E_{0}^{2}a}\;\equiv\; \dfrac{\xi_\text{UV}^{(1)}}{a}\;.
\label{xiUVIR}
\end{eqnarray}
$\xi_\text{IR}^{(1)}$ and $\xi_\text{UV}^{(1)}$ are respectively the values of 
$\xi_\text{IR}(a)$ and $\xi_\text{UV}(a)$ at $a=1$.
They meet at $a=\sqrt{(\mu_\text{UV}E_0)/(\mu_\text{IR}E_\text{Planck})}=\sqrt{\xi_\text{UV}^{(1)}/\xi_\text{IR}^{(1)}}$.
This behaviors of $\xi_\text{IR}(a)$ and $\xi_\text{UV}(a)$ for the case $\xi_\text{IR}^{(1)}=1$, $\xi_\text{UV}^{(1)}=3$ is shown in Figure.~\ref{cutoffplot}. 

\begin{figure}[ht]
\hspace{-1cm}
\subfigure[$\xi_\text{IR}^{(1)}=2$]{\includegraphics[width=7.5cm]{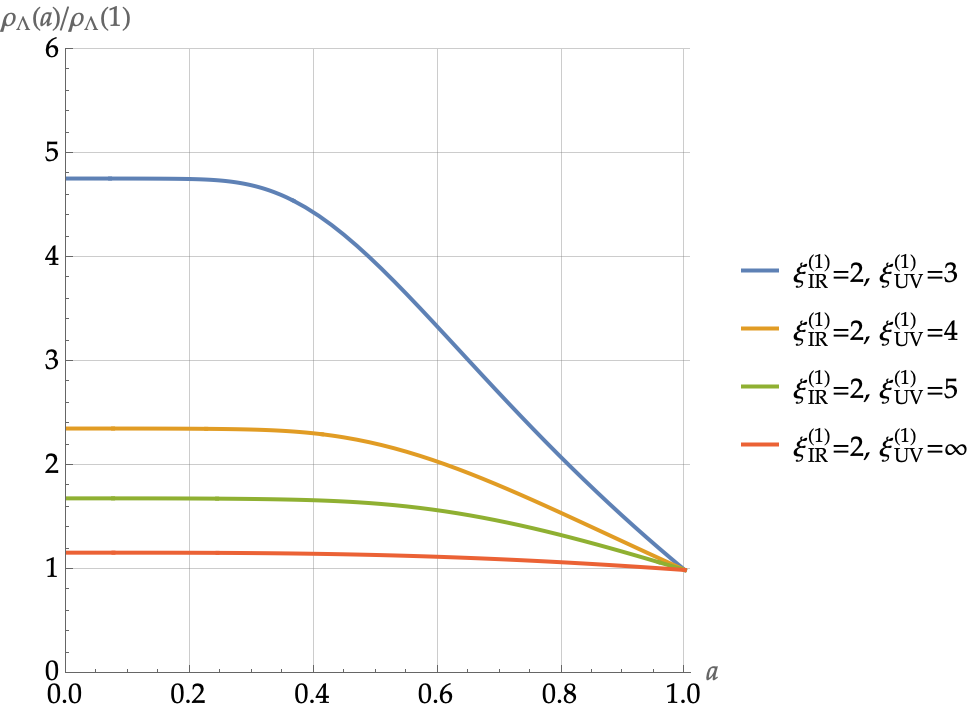}}
\subfigure[$\xi_\text{UV}^{(1)}=5$]{\includegraphics[width=7.5cm]{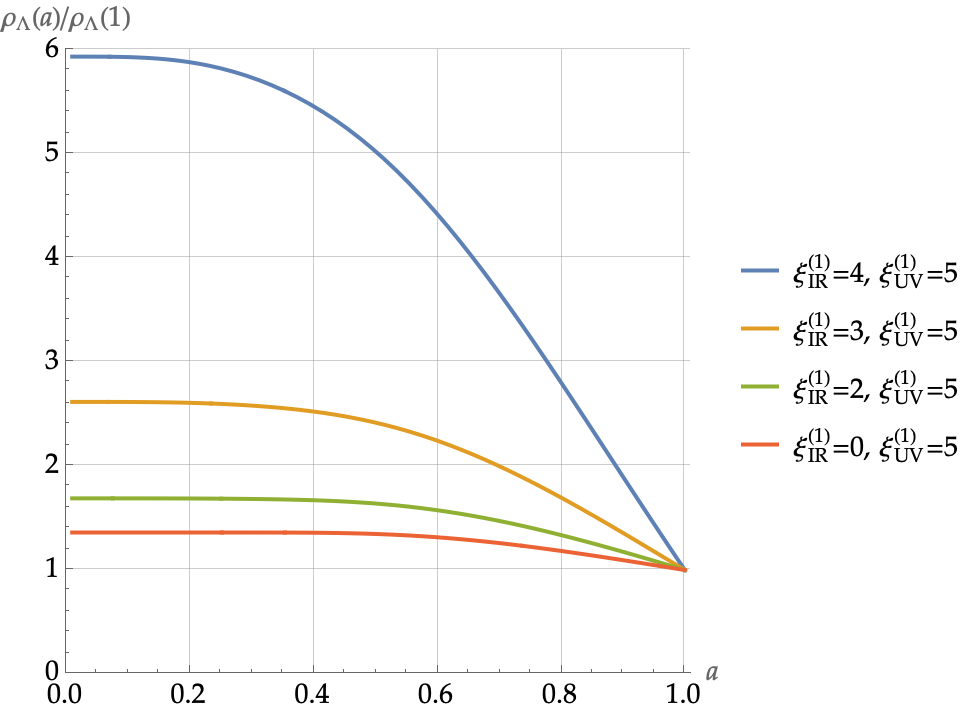}}
\caption{$\rho_\Lambda(a)/\rho_\Lambda(1)$ for 
several values of $\xi_\text{IR}^{(1)}$ and $\xi_\text{UV}^{(1)}$.
The $(\xi_\text{IR}^{(1)},\xi_\text{UV}^{(1)})=(2,5)$ case is shown on both graphs.
$\rho_\Lambda(a)$ approaches a constant as $(\xi_\text{IR}^{(1)},\xi_\text{UV}^{(1)})\to(0,\infty)$.
}
\label{rhoUVIRplot0}
\end{figure}

\begin{figure}[ht]
\hspace{-1cm}
\subfigure[$\xi_\text{IR}^{(1)}=2$]{\includegraphics[width=7.5cm]{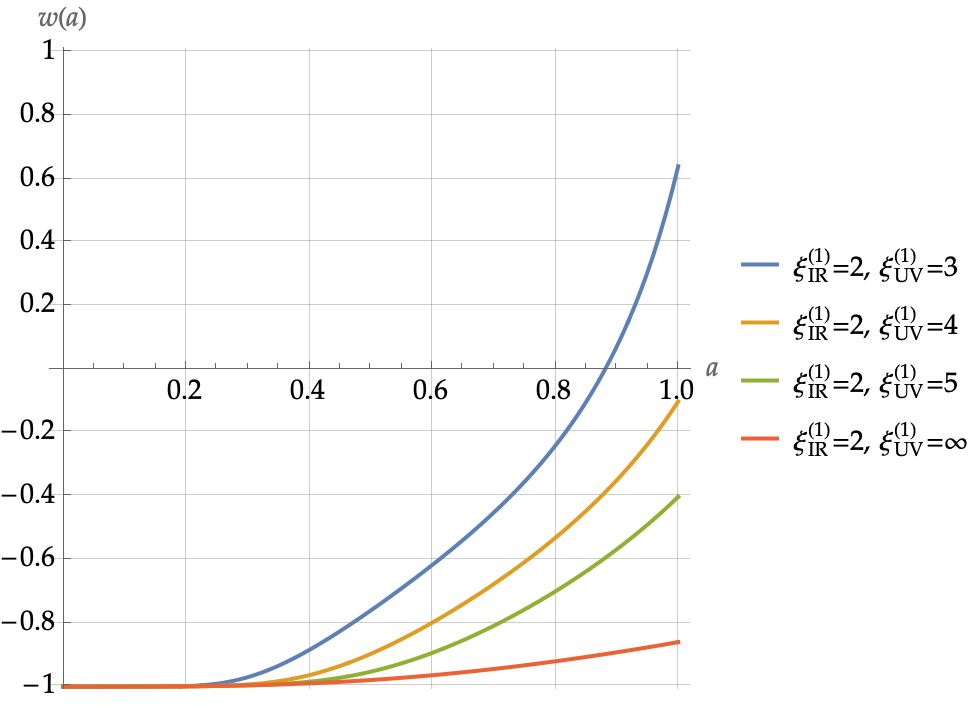}}
\subfigure[$\xi_\text{UV}^{(1)}=5$]{\includegraphics[width=7.5cm]{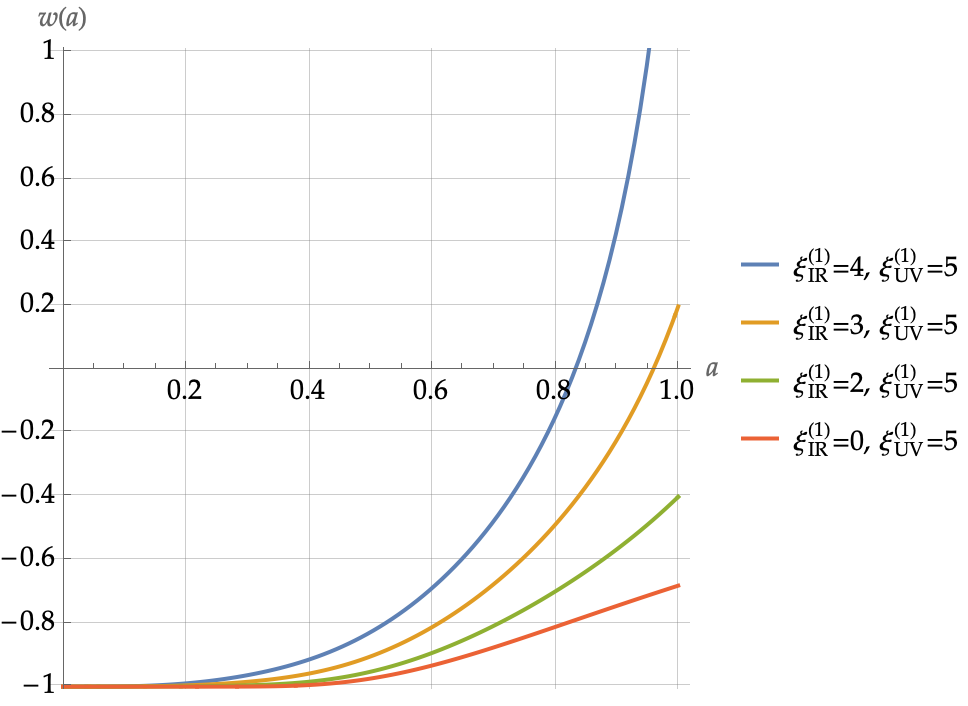}}
\caption{$w(a)$ for 
several values of $\xi_\text{IR}^{(1)}$ and $\xi_\text{UV}^{(1)}$.
The $(\xi_\text{IR}^{(1)},\xi_\text{UV}^{(1)})=(2,5)$ case is shown on both graphs.
Note that $w(a)\ge -1$ in all cases, and
$w(a)\to -1$ as $(\xi_\text{IR}^{(1)},\xi_\text{UV}^{(1)})\to(0,\infty)$
for all $a$.
The slope of $w(a)$ at $a=1$ is always positive,
which means that $w_a=-w'(1)$ is always negative.
}
\label{wUVIRplot0}
\end{figure}

Using~\eqref{xiUVIR},
the dark energy density is given by~\eqref{eq:Lofz}, and the equation-of-state parameter is given by~\eqref{wageneric}.
Since
\begin{equation}
\xi_\text{IR}(a)\xrightarrow{a\to 0} 0\;,\qquad
\xi_\text{UV}(a)\xrightarrow{a\to 0} \infty\;,
\end{equation}
we have
\begin{equation}
b(\xi_\text{IR}(a))\xrightarrow{a\to 0} 1\;,\qquad
b(\xi_\text{UV}(a))\xrightarrow{a\to 0} 0\;,
\end{equation}
and
\begin{equation}
\rho_\Lambda(0)/\rho_* \;=\; 1\;,\qquad
w(0) \;=\; -1\;.
\end{equation}
In the limits $\xi_\text{IR}^{(1)}\to 0$
and $\xi_\text{UV}^{(1)}\to\infty$, the expressions simplify to
\begin{eqnarray}
\rho_\Lambda(a)/\rho_*
& & \begin{cases}
\xrightarrow{\xi_\text{IR}^{(1)}\to 0} & 
1 - b(\xi_\text{UV}) 
\;,\\
\xrightarrow{\xi_\text{UV}^{(1)}\to \infty} &
b(\xi_\text{IR})\;,
\end{cases}
\\
& & \vphantom{x}\cr
w(a)
& & \begin{cases} 
\xrightarrow{\xi_\text{IR}^{(1)}\to 0} & 
-1 -\dfrac{1}{18\big[1-b(\xi_\text{UV})\big]}
\bigg[
 \xi_\text{UV}^3\, e^{-\xi_\text{UV}}\bigg( a\,\dfrac{d\xi_\text{UV}}{da} \bigg)
\bigg]
\;,
\\
\xrightarrow{\xi_\text{UV}^{(1)}\to \infty} &-1 +\dfrac{1}{18\,b(\xi_\text{IR})}
\bigg[
\xi_\text{IR}^3\, e^{-\xi_\text{IR}}\bigg( a\,\dfrac{d\xi_\text{IR}}{da} \bigg)
\bigg]
\;.
\end{cases}
\end{eqnarray}
When both $\xi_\text{IR}^{(1)}\to 0$ and $\xi_\text{UV}^{(1)}\to\infty$,
we have $\rho_\Lambda(a)=1$ and $w(a)=-1$.
The behaviors of $\rho_\Lambda(a)$ and $w(a)$ are shown for several values of 
$(\xi_\text{IR}^{(1)},\xi_\text{UV}^{(1)})$ in
Figures~\ref{rhoUVIRplot0} and~\ref{wUVIRplot0}.

\begin{figure}[ht]
\centering
\hspace{-0.5cm}
\subfigure[]{\includegraphics[height=6.5cm]{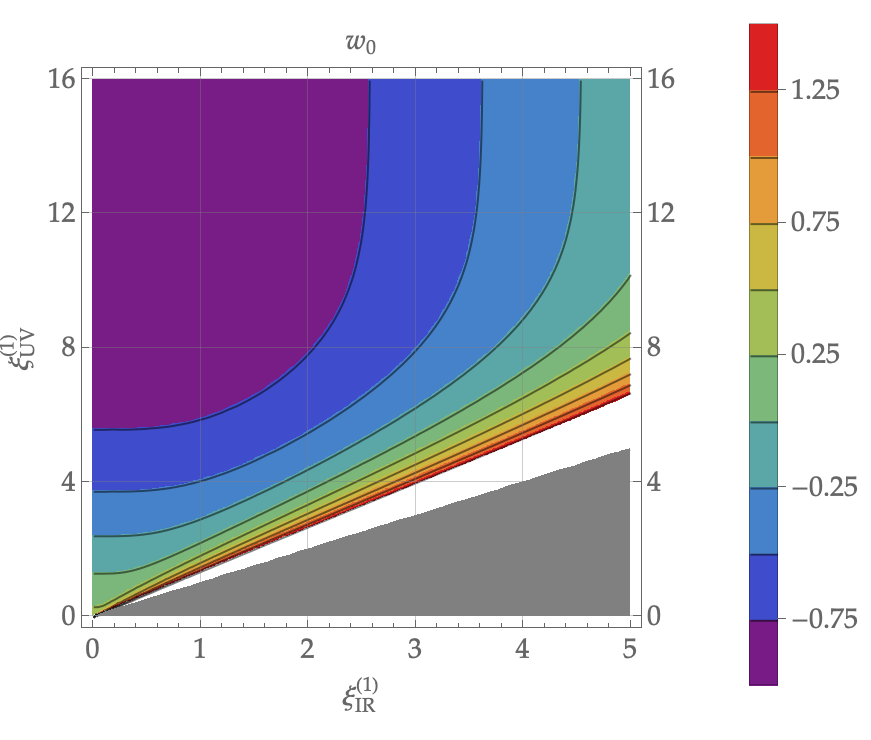}}
\hspace{0.5cm}
\subfigure[]{\includegraphics[height=6.5cm]{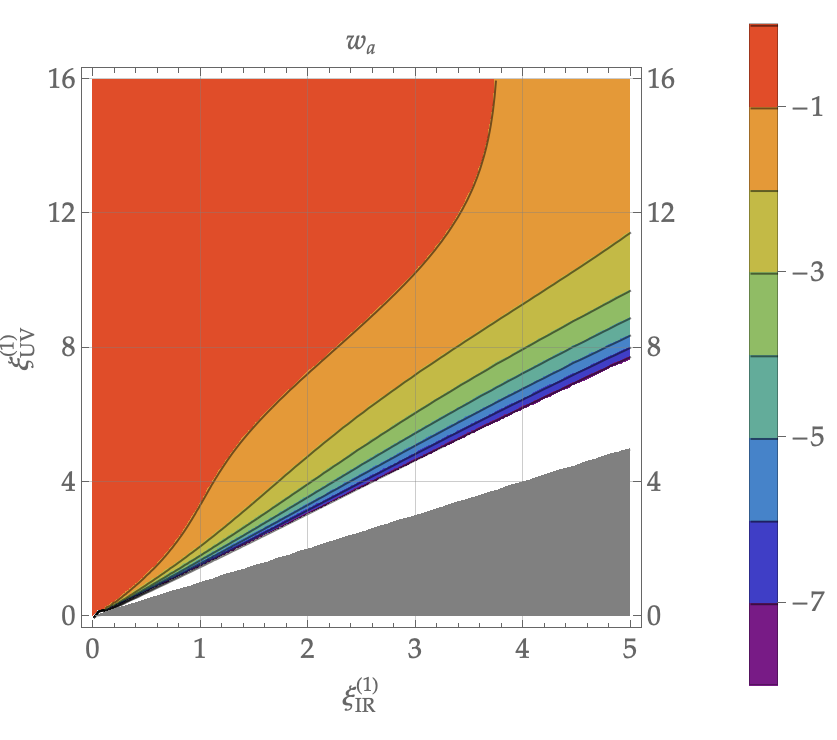}}
\caption{Contour plots of
(a) $w_0(\xi_\text{IR}^{(1)},\xi_\text{UV}^{(1)})$, and
(b) $w_a(\xi_\text{IR}^{(1)},\xi_\text{UV}^{(1)})$.
The gray regions in both plots where $\xi_\text{IR}^{(1)}>\xi_\text{UV}^{(1)}$ are forbidden.  The white region in (a) is where
$w_0(\xi_\text{IR}^{(1)},\xi_\text{UV}^{(1)})>1.5$ and the white region in (b) is where $w_a(\xi_\text{IR}^{(1)},\xi_\text{UV}^{(1)})<-8$.
As the boundary with the gray area where $\xi_\text{IR}^{(1)}=\xi_\text{UV}^{(1)}$ is approached,
$w_0(\xi_\text{IR}^{(1)},\xi_\text{UV}^{(1)})\to\infty$ and $w_a(\xi_\text{IR}^{(1)},\xi_\text{UV}^{(1)})\to -\infty$.
}
\label{w0waUVIRcontourplot}
\end{figure}

\begin{figure}[ht]
\centering
\hspace{-0.5cm}
\subfigure[]{\includegraphics[width=7cm]{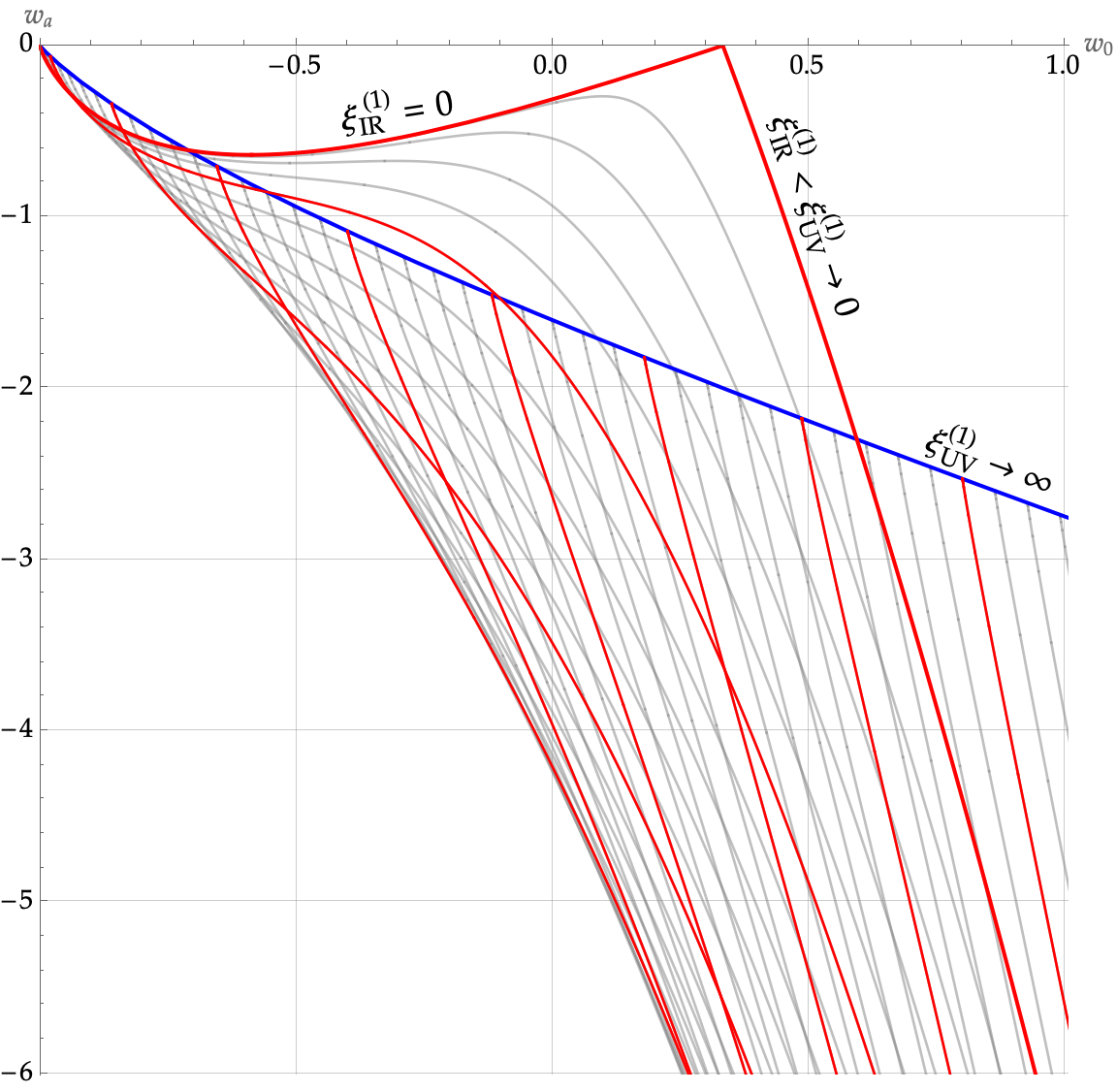}}
\hspace{0.5cm}
\subfigure[]{\includegraphics[width=7cm]{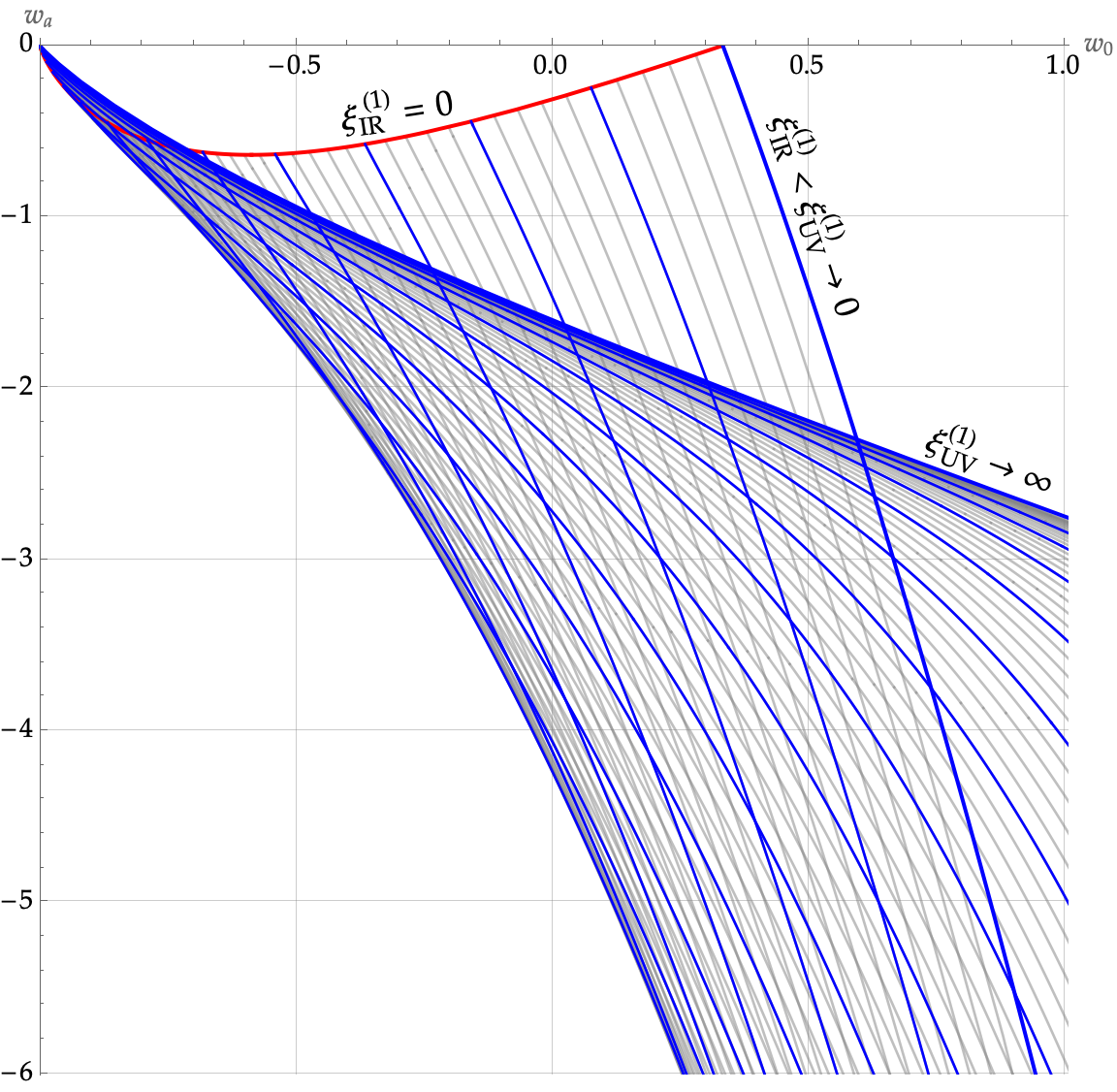}}
\caption{$(w_0,w_a)$ for various values of $(\xi_\text{IR}^{(1)},\xi_\text{UV}^{(1)})$.
(a) The gray and red lines connect points which share the same value of $\xi_\text{IR}^{(1)}$, starting from $\xi_\text{IR}^{(1)}=0$
at 0.2 intervals, with the red lines corresponding to integer values of $\xi_\text{IR}^{(1)}$.
The $\xi_\text{IR}^{(1)}=0$ line has a kink at 
$(w_0,w_a)=(\frac{1}{3},0)$.  
The $w_0<1/3$ part of the line is given by~\eqref{w0waUV}, which is reached if $\xi_\text{UV}^{(1)}$ is kept constant as
$\xi_\text{IR}^{(1)}\to 0$.
The $w_0>1/3$ part of the line is given by~\eqref{w0wakappaLimit}, which is reached 
if the ratio $\xi_\text{UV}^{(1)}/\xi_\text{IR}^{(1)}$ is kept constant as $\xi_\text{IR}^{(1)}\to 0$.
(b) The gray and blue lines connect points which share the same value of $\xi_\text{UV}^{(1)}$, starting from $\xi_\text{UV}^{(1)}=0$
at 0.2 intervals, with the blue lines corresponding to integer values of $\xi_\text{UV}^{(1)}$.
The $\xi_\text{UV}^{(1)}=0$ line is also the $w_0>\frac{1}{3}$ part of the
$\xi_\text{IR}^{(1)}=0$ line, since we must maintain
$\xi_\text{IR}^{(1)}<\xi_\text{UV}^{(1)}$ as $\xi_\text{UV}^{(1)}\to 0$. 
The $\xi_\text{UV}^{(1)}\to\infty$ line is given by~\eqref{w0waIR}.
}
\label{w0waUVIRplot}
\end{figure}

\noindent
What is most important, $w_{0}=w(1)$ and $w_a=-w'(1)$ are given by
\begin{eqnarray}
w_0 \;=\; w(1) 
& = &
-1 + \dfrac{
\big(\xi_\text{UV}^{(1)}\big)^4 e^{-\xi_\text{UV}^{(1)}}+
\big(\xi_\text{IR}^{(1)}\big)^4 e^{-\xi_\text{IR}^{(1)}}
}
{18\Big[
b\big(\xi_\text{IR}^{(1)}\big)-
b\big(\xi_\text{UV}^{(1)}\big)
\Big]}  
\vphantom{\Bigg|}\\
& \xrightarrow{\xi_\text{IR}^{(1)}\to 0} &
-1 + \dfrac{
\big(\xi_\text{UV}^{(1)}\big)^4 e^{-\xi_\text{UV}^{(1)}}
}
{18\Big[
1-
b\big(\xi_\text{UV}^{(1)}\big)
\Big]}  
\;,\vphantom{\Bigg|}\cr
& \xrightarrow{\xi_\text{UV}^{(1)}\to\infty} &
-1 + \dfrac{
\big(\xi_\text{IR}^{(1)}\big)^4 e^{-\xi_\text{IR}^{(1)}}
}
{18\,
b\big(\xi_\text{IR}^{(1)}\big)
}
\;=\; -1 + \dfrac{
\big(\xi_\text{IR}^{(1)}\big)^4
}
{\left[1
+\big(\xi_\text{IR}^{(1)}\big)
+\frac{1}{2}\big(\xi_\text{IR}^{(1)}\big)^2
+\frac{1}{6}\big(\xi_\text{IR}^{(1)}\big)^3
\right]
}
\;,\vphantom{\Bigg|}\cr
w_a \;=\; -w'(1)
& = & 
\frac{
\big(4-\xi_\text{UV}^{(1)}\big)\big(\xi_\text{UV}^{(1)}\big)^4 e^{-\xi_\text{UV}^{(1)}}-
\big(4-\xi_\text{IR}^{(1)}\big)\big(\xi_\text{IR}^{(1)}\big)^4 e^{-\xi_\text{IR}^{(1)}}
}
{18\Big[b(\xi_\text{IR}^{(1)})-b(\xi_\text{UV}^{(1)})\Big]} 
\vphantom{\Bigg|}\cr
& & 
-\,3\left[
\dfrac{
\big(\xi_\text{UV}^{(1)}\big)^4 e^{-\xi_\text{UV}^{(1)}}+
\big(\xi_\text{IR}^{(1)}\big)^4 e^{-\xi_\text{IR}^{(1)}}
}
{18\Big[b(\xi_\text{IR}^{(1)})-b(\xi_\text{UV}^{(1)})\Big]}  
\right]^2
\vphantom{\Bigg|}\\
& = &
\Bigg[
\dfrac{
\big(4-\xi_\text{UV}^{(1)}\big)\big(\xi_\text{UV}^{(1)}\big)^4 e^{-\xi_\text{UV}^{(1)}}-
\big(4-\xi_\text{IR}^{(1)}\big)\big(\xi_\text{IR}^{(1)}\big)^4 e^{-\xi_\text{IR}^{(1)}}
}
{
\big(\xi_\text{UV}^{(1)}\big)^4 e^{-\xi_\text{UV}^{(1)}}+
\big(\xi_\text{IR}^{(1)}\big)^4 e^{-\xi_\text{IR}^{(1)}}
}
\Bigg]
(w_0+1)
-3(w_0+1)^2
\vphantom{\Bigg|}\cr
& \xrightarrow{\xi_\text{IR}^{(1)}\to 0} &
\big(4-\xi_\text{UV}^{(1)}\big)(w_0+1)
-3(w_0+1)^2
\;,\vphantom{\Bigg|}
\label{w0waUV}
\\
& \xrightarrow{\xi_\text{UV}^{(1)}\to\infty} &
-\big(4-\xi_\text{IR}^{(1)}\big)(w_0+1)
-3(w_0+1)^2
\;.\vphantom{\Bigg|}
\label{w0waIR}
\end{eqnarray}
Compare with Eq.~\eqref{w0wa}.
If we use the ratio $\kappa = \xi_\text{UV}^{(1)}/\xi_\text{IR}^{(1)} > 1$ as a parameter, we find
\begin{eqnarray}
w_0 & = & -1 + \dfrac{\big(\xi_\text{IR}^{(1)}\big)^4
\Big[
\kappa^4 e^{-\kappa\xi_\text{IR}^{(1)}}+
e^{-\xi_\text{IR}^{(1)}}
\Big]
}
{18\Big[
b\big(\xi_\text{IR}^{(1)}\big)-
b\big(\kappa\xi_\text{IR}^{(1)}\big)
\Big]}  
\cr
& \xrightarrow{\xi_\text{IR}^{(1)}\to 0} &
-1+\dfrac{4(\kappa^4+1)}{3(\kappa^4-1)}\;=\;
\dfrac{\kappa^4+7}{3(\kappa^4-1)} 
\;\xrightarrow{\kappa\to\infty}\; \dfrac{1}{3}
\;,
\vphantom{\Bigg|}\cr
w_a
& = & 
\Bigg[
\dfrac{
\big(4-\kappa\xi_\text{IR}^{(1)}\big)\big(\kappa \xi_\text{IR}^{(1)}\big)^4 e^{-\kappa\xi_\text{IR}^{(1)}}-
\big(4-\xi_\text{IR}^{(1)}\big)\big(\xi_\text{IR}^{(1)}\big)^4 e^{-\xi_\text{IR}^{(1)}}
}
{
\big(\kappa\xi_\text{IR}^{(1)}\big)^4 e^{-\kappa\xi_\text{IR}^{(1)}}+
\big(\xi_\text{IR}^{(1)}\big)^4 e^{-\xi_\text{IR}^{(1)}}
}
\Bigg]
(w_0+1)
-3(w_0+1)^2
\vphantom{\Bigg|}\cr
& \xrightarrow{\xi_\text{IR}^{(1)}\to 0} &
\dfrac{4(\kappa^4-1)}{\kappa^4+1}(w_0+1)
-3(w_0+1)^2
\;=\; \dfrac{16}{3}-3(w_0+1)^2
\;=\; -\dfrac{64\kappa^4}{3(\kappa^4-1)^2}
\;\xrightarrow{\kappa\to\infty}\;
0 
\;.\vphantom{\Bigg|}
\label{w0wakappaLimit}
\cr
& &
\end{eqnarray}
The values of 
$(w_0,w_a)$ for various choices of $(\xi_\text{IR}^{(1)},\xi_\text{UV}^{(1)})$
are shown in Figures~\ref{w0waUVIRcontourplot} and~\ref{w0waUVIRplot}.

\begin{figure}[ht]
\centering
\includegraphics[height=7cm]{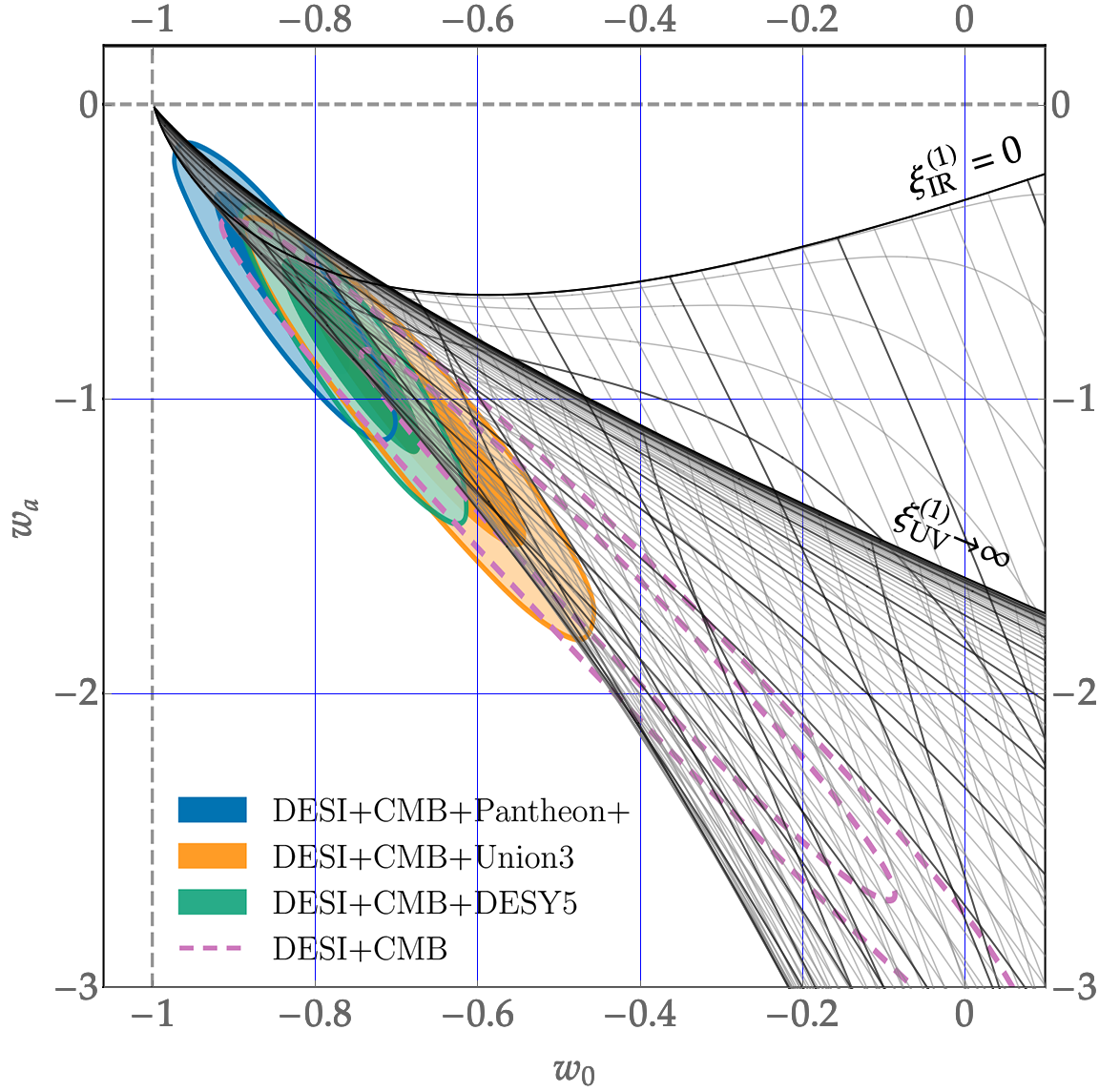}
\caption{
The equation-of-state parameter $w$ as a function of the scale factor $a$ is parametrized as $w(a)=w_0+w_a(1-a)$ around $a=1$.
The values of $(w_0,w_a)$ for various values of
$(\xi_\text{IR}^{(1)},\xi_\text{UV}^{(1)})$ 
compared against the 68\% and 95\% likelihood contours from DESI, taken from Figure~11 of Ref.~\cite{DESI:2025zgx}. 
The thin black lines connect the points that share the same value of 
$\xi_\text{IR}^{(1)}$ or $\xi_\text{UV}^{(1)}$.
The lines are plotted for values of $\xi_\text{IR}^{(1)}$ and $\xi_\text{UV}^{(1)}$ at $0.2$ intervals as in Figure~\ref{w0waUVIRplot}.
}
\label{ComparisonWithDESI}
\end{figure}

\begin{figure}[ht]
\centering
\subfigure[]{\includegraphics[height=7cm]{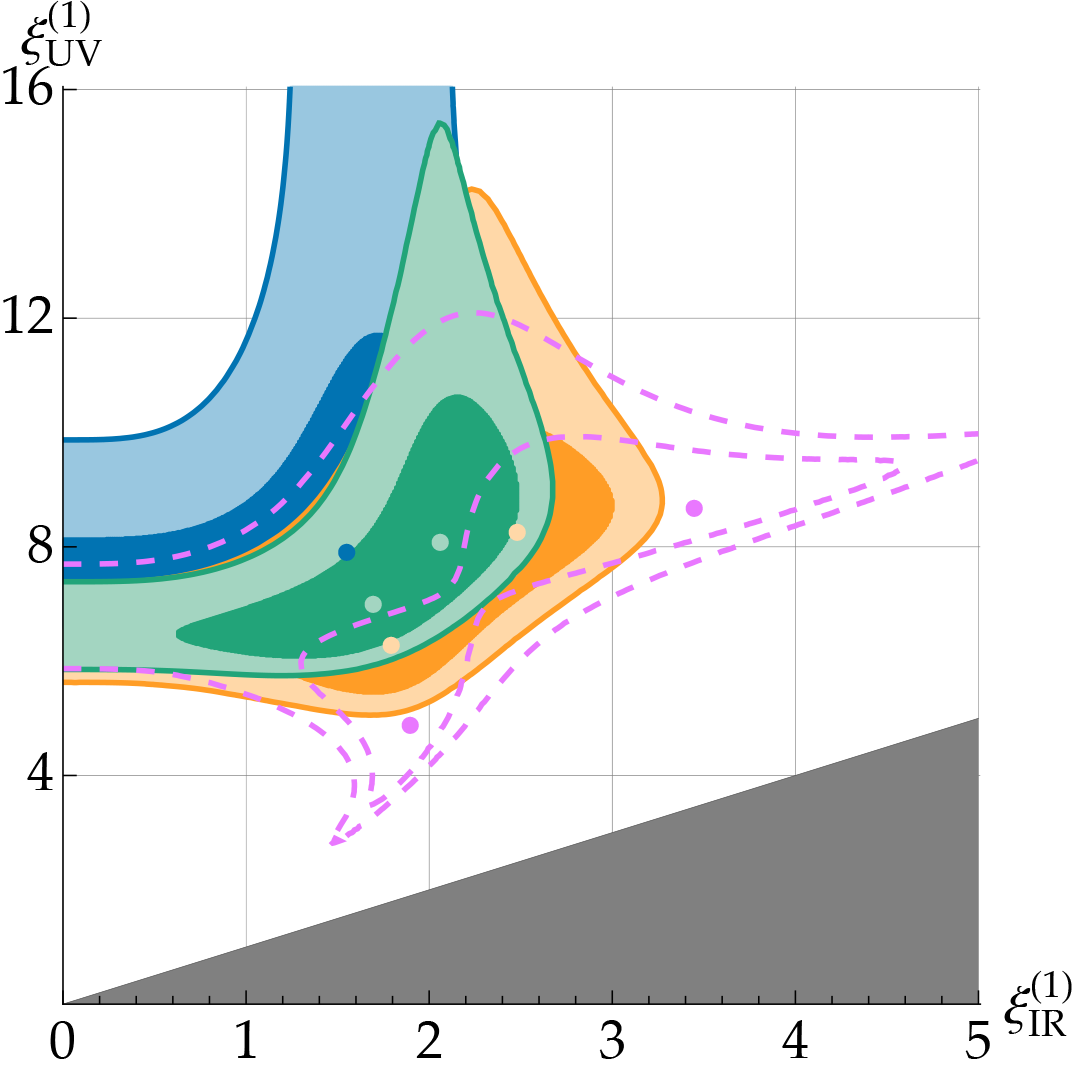}}
\subfigure[]{\includegraphics[height=7.2cm]{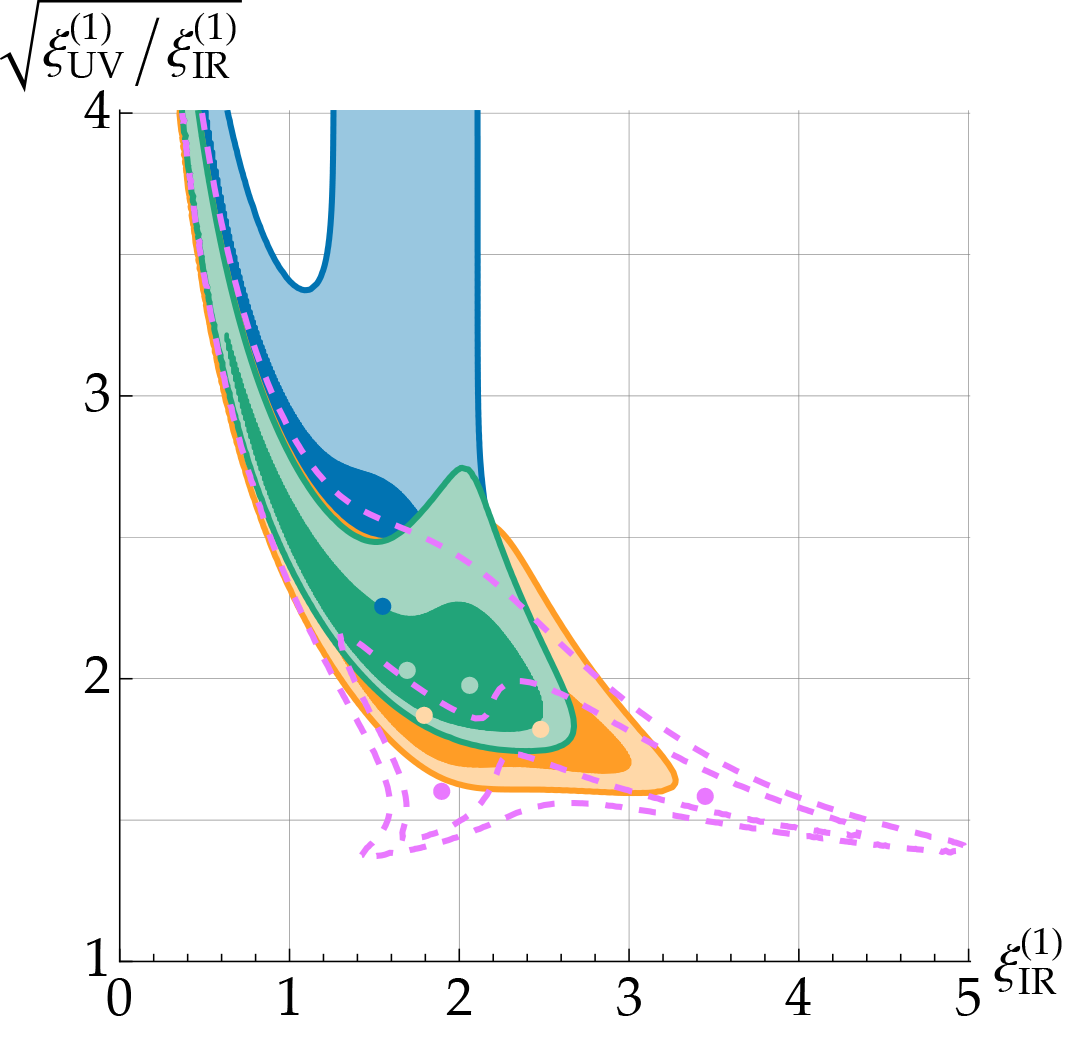}}
\caption{(a) The regions of $(\xi_\text{IR}^{(1)},\xi_\text{UV}^{(1)})$ preferred by DESI.
The color coding of the likelihood contours is the same as in Figure~\ref{ComparisonWithDESI}.
The colored circles indicate the maximum likelihood points for each case. Note that the values of $(\xi_\text{IR}^{(1)},\xi_\text{UV}^{(1)})$ that correspond to
specific values of $(w_0,w_a)$ 
do not necessarily exist nor are they unique.
The shaded area is forbidden since we must have
$\xi_\text{IR}^{(1)}<\xi_\text{UV}^{(1)}$.
(b) The likelihood contours transferred to the $(\xi_\text{IR}^{(1)},\sqrt{\xi_\text{UV}^{(1)}/\xi_\text{IR}^{(1)}})$ plane.
The value of $\sqrt{\xi_\text{UV}^{(1)}/\xi_\text{IR}^{(1)}}$ gives the
scale $a$ at which the UV and IR cutoffs come together in this model.
}
\label{ComparisonWithDESI2}
\end{figure}

In Figure~\ref{ComparisonWithDESI}, the possible values
of $(w_0,w_a)$ in this model are compared to the constraints from DESI. 
The likelihood contours are taken from Figure~11 of Ref.~\cite{DESI:2025zgx}.
They overlap with the regions of $(w_0,w_a)$ that can be
accommodated in this model, thereby providing constraints
on the model parameters $(\xi_\text{IR}^{(1)},\xi_\text{UV}^{(1)})$.

The likelihood contours transferred to the $(\xi_\text{IR}^{(1)},\xi_\text{UV}^{(1)})$ plane are shown in
Figure~\ref{ComparisonWithDESI2}(a).
Figure~\ref{ComparisonWithDESI2}(b) shows the contours
converted to those for $\xi_\text{IR}^{(1)}$ and $\sqrt{\xi_\text{UV}^{(1)}/\xi_\text{IR}^{(1)}}$, the later of which gives the value of the scale parameter $a$ at which the UV and IR cutoffs come together in the future in this model.

Note that the values of $(\xi_\text{IR}^{(1)},\xi_\text{UV}^{(1)})$ that 
correspond to given values of $(w_0,w_a)$ are found by looking
for the intersections of the corresponding lines on the coutour plots
shown in Figures~\ref{w0waUVIRcontourplot}(a) and (b).
These curves may or may not intersect, and if they do, may have multiple intersections.
Consequently, each point on the $(w_0,w_a)$ plane 
may correspond to no, one, or more
points on the $(\xi_\text{IR}^{(1)},\xi_\text{UV}^{(1)})$ plane.
Indeed, the maximum likelihood points for the four cases
listed in Figure~\ref{ComparisonWithDESI}(a) correspond to the
following points on $(\xi_\text{IR}^{(1)},\xi_\text{UV}^{(1)})$:
\begin{itemize}
\item \textbf{DESI+CMB} (Eq.~(25) of Ref.~\cite{DESI:2025zgx})
\begin{equation}
(w_0,w_a)\,=\,(-0.42,-1.75) \quad\to\quad
(\xi_\text{IR}^{(1)},\xi_\text{UV}^{(1)})
\,=\,(1.90,4.90)\,,\;(3.45.8.71)\,.
\end{equation}

\item \textbf{DESI+CMB+Pantheon+} (Eq.~(26) of Ref.~\cite{DESI:2025zgx})
\begin{equation}
(w_0,w_a)\,=\,(-0.838,-0.62) \quad\to\quad
(\xi_\text{IR}^{(1)},\xi_\text{UV}^{(1)})
\,=\,(1.55,7.93)\,.
\end{equation}

\item \textbf{DESI+CMB+Union3} (Eq.~(27) of Ref.~\cite{DESI:2025zgx})
\begin{equation}
(w_0,w_a)\,=\,(-0.667,-1.09) \quad\to\quad
(\xi_\text{IR}^{(1)},\xi_\text{UV}^{(1)})
\,=\,(1.79,6.31)\,,\;(2.48.8.28)\,.
\end{equation}

\item \textbf{DESI+CMB+DESY5} (Eq.~(28) of Ref.~\cite{DESI:2025zgx})
\begin{equation}
(w_0,w_a)\,=\,(-0.752,-0.86) \quad\to\quad
(\xi_\text{IR}^{(1)},\xi_\text{UV}^{(1)})
\,=\,(1.69,7.02)\,,\;(2.06.8.10)\,.
\end{equation}

\end{itemize}
These maximum likelihood points are indicated by colored circles on Figure~\ref{ComparisonWithDESI2}.
The existence of multiple preferred points in 
$(\xi_\text{IR}^{(1)},\xi_\text{UV}^{(1)})$ leads to the
double-blob structure which is clearly visible in 
the 68\% likelihood contour for the DESI+CMB case.

\noindent
\section{Discussion}\label{sec:discussion}

In this paper, we have discussed some consequences of an intrinsically non-commutative and T-duality covariant formulation of string theory, the metastring formulation, for dark energy .
Using general precepts of spacetime non-locality and Lorentz covariance in quantum gravity, we are led inexorably to infinite statistics to describe the statistics of distinguishable quanta.
Such statistics are naturally realized in the metastring formulation of string theory.
Associating these quanta with dark energy, we propose a density of states that follows the Wien distribution associated with infinite statistics.
(Note that our computation could be generalized for other distributions as well.)
In this approach, dark energy originates from the curvature of dual spacetime in the metastring formulation, which upon quantization, is made out of ``quanta,'' that obey infinite statistics. 
Infinite statistics, plus a particular form the relation between the UV and IR physics leads us to demonstrate quite generically, that the dark energy %
is dynamical, \textit{i.e.}, time-dependent, and that its effective equation-of-state parameter satisfies $w(a) > -1$. This raises intriguing questions about the dynamical nature of fundamental constants in the context of a full theory of quantum gravity and establishes the central role of T-duality and UV/IR relations in the nature of dark energy. 
Within this framework, we compute $w_0 = w(1)$ and $w_a = -w'(1)$ and favorably compare our results to recent observations by DESI as summarized in Figure~\ref{ComparisonWithDESI2}.
It would be interesting to understand what our results imply for the Hubble tension~\cite{DiValentino:2021izs} in light of~\cite{DESI:2025zgx} (\textit{cf.}, Figure~9 therein).
Given our proposal for the origin of dark energy from the geometry of the dual spacetime, the latest results from DESI might point to a fundamentally new understanding of quantum spacetime in the context of quantum gravity~\cite{Hubsch:2024agh}.

\acknowledgments

We thank 
P.~Berglund, D.~Edmonds, {T.~Eifler}, J.~Erlich, L.~Freidel, A.~Geraci, S.~Horiuchi, T.~H\"ubsch, J.~Kowalski-Glikman, R.~G.~Leigh, D.~Mattingly and J.~Papa for helpful discussions. 
SH, DM, and TT are supported in part by the U.S.\ Department of Energy (DE-SC0020262, Task B) and the Julian Schwinger Foundation.
VJ is supported in part by the South African Research Chairs Initiative of the National Research Foundation, grant number 78554, and is grateful to the African Institute for Mathematical Sciences (AIMS) Ghana for their hospitality during the completion of this work.
MJK is supported by the National Science Foundation (AST-2011731).

\appendix

\section{Derivation of the equation-of-state parameter}
\label{app:w}

In this appendix we review the derivation of the 
time-dependence of the equation-of-state parameter $w$
from the time-dependence of the dark energy density.
Here, we use the scale parameter $a$ as a proxy for the redshift
$z=\frac{1}{a}-1$ to facilitate the derivation of 
\begin{equation}
w_a = -\dfrac{dw}{da}\bigg|_{a=1}
\end{equation}
later.

In a matter-dominated flat universe, the Friedmann equation tells us that
\begin{equation}
H(a)^2 
\;=\; \bigg(\dfrac{\dot{a}}{a}\bigg)^2
\;=\;
\frac{8\pi G_N}{3} \sum_i \rho_i(a) \;\approx\; \frac{8\pi G_N}{3} \Big[\, \rho_m(a) + \rho_\Lambda(a) \,\Big] \;.
\end{equation}
Rearranging terms, we have
\begin{equation}
\rho_\Lambda(a) 
\;=\; \frac{3H(a)^2}{8\pi G_N} - \rho_m(a)
\;=\; \frac{3H(a)^2}{8\pi G_N} \Big[ 1 - \Omega_m(a) \Big] \;,
\end{equation}
where
\begin{equation}
\Omega_i(a) \;=\; \dfrac{8\pi G_N}{3H(a)^2}\,\rho_i(a)\;.
\end{equation}
Next, the deceleration parameter is
\begin{equation}
q(a) \;=\; -\frac{\ddot{a}}{a H(a)^2} \;=\; \frac{4\pi G_N}{3 H(a)^2} \sum_i \Big[ \rho_i(a) + 3 p_i(a) \Big] \;\approx\; \frac{1}{2} + \frac{4\pi G_N}{H(a)^2}\,p_\Lambda(a) \;,
\end{equation}
the factor of $\frac{1}{2}$ coming from the contribution
of $\sum_i\rho_i\approx\rho_m + \rho_\Lambda$.
Thus,
\begin{equation}
p_\Lambda(a) \;=\; \frac{H(a)^2}{4\pi G_N} \bigg[ q(a) - \frac{1}{2} \bigg] \;.
\end{equation}
The ratio of pressure to energy density gives the effective equation-of-state parameter for dark energy:
\begin{equation}
w(a)
\;=\; \frac{p_\Lambda(a)}{\rho_\Lambda(a)} 
\;=\; \frac{2 q(a) - 1}{3\big[1-\Omega_m(a)\big]} \;.
\label{eq:poverrho}
\end{equation}
To simplify the denominator of \eqref{eq:poverrho}, we first note that
\begin{equation}
\Omega_m(a) \;=\; \bigg[\frac{H_0}{H(a)}\bigg]^2 \dfrac{\Omega_{m,0}}{a^3} \;.
\end{equation}
where $\Omega_{m,0}$ is the value of $\Omega_m(a)$ today ($a=1$, $z=0$).
Since
\begin{equation}
H(a) \;=\; H_0 \sqrt{ \dfrac{\Omega_{m,0}}{a^{3}} + \Omega_\Lambda(a) } \;,
\label{Ha}
\end{equation}
we can write
\begin{equation}
3\Big[1-\Omega_m(a)\Big] \;=\; \dfrac{3H_0^2}{H(a)^2}\,\Omega_\Lambda(a)\;.
\label{denom}
\end{equation}
To obtain an expression for the numerator of \eqref{eq:poverrho}, we then observe that
\begin{equation}
\frac{dH}{dt} \;=\; \frac{\ddot{a}}{a} - \left( \frac{\dot{a}}{a} \right)^2 \;=\; -H(a)^2 \Big[ q(a) + 1 \Big] \;,
\end{equation}
which means
\begin{equation}
q(a) \;=\; -\frac{1}{H(a)^2} \frac{dH(a)}{dt} - 1 \;=\; 
-\dfrac{a}{H(a)}\dfrac{dH(a)}{da} - 1 \;,
\end{equation}
Then using~\eqref{Ha} again, we find after some straightforward algebra that
\begin{equation}
2q(a)-1 \;=\;
 \frac{3H_0^2}{H(a)^2} \bigg[ 
-\frac{a}{3}\,\frac{d\Omega_\Lambda(a)}{da} - \Omega_\Lambda(a) 
\bigg]
\label{num}
\end{equation}
Putting~\eqref{denom} and \eqref{num} together, we obtain
\begin{equation}
w(a) \;=\; \frac{-\dfrac{a}{3}\,\dfrac{d\Omega_\Lambda(a)}{da} - \Omega_\Lambda(a)}{\Omega_\Lambda(a)}
\;=\; -1 - \frac{a}{3}\,\frac{d}{da} \log \Omega_\Lambda(a) \;.
\label{wafromOmegaLambda}
\end{equation}
If $\Lambda(a) = \text{constant}$, we verify that $w=-1$ as required.

\section{Quantum spacetime and quantum gravity}\label{sec:qsqg}

In this appendix, we outline the non-perturbative formulation of
quantum gravity in terms of a doubled matrix model quantum theory proposed in~\cite{Freidel:2013zga, Freidel:2015uug, Freidel:2016pls, Freidel:2017xsi, Freidel:2017wst}, \textit{i.e.}, the metastring.
In this description, everything is built out of partonic degrees of freedom 
represented by the entries of the quantum gravitational matrix model, and, 
in the leading term in the expansion involving the fundamental length, 
dark energy is realized as a dynamical geometry of dual spacetime.

The starting point of the metastring formalism is the following worldsheet action~\cite{Tseytlin:1990nb,Tseytlin:1990va},
which is chiral, doubles the degrees of freedom (\textit{i.e.}, works in phase space), and is manifestly
invariant under Born reciprocity/T-duality:
\be
S_{2d}\;{=}\;\frac{1}{4\pi}\int_{\Sigma}
\Bigl[\,
  \partial_{\tau}   \X^{A} (\eta_{AB}(\X)+\omega_{AB}(\X))
- \partial_{\sigma} \X^{A} \,H_{\!AB} (\X)
\,\Bigr] \partial_{\sigma} \X^{B}
\;.
\label{e:MSA}
\ee 
Here $\Sigma$ is the worldsheet, the doubled target space variables 
$\X^A = (x^a/\lambda,\tilde{x}_a/\lambda)$ combine 
the sum ($x = x_L+x_R$) and the difference ($\tx = x_L-x_R$) of the left- and right-movers on the string
($a,A=0,1,\cdots,d-1=25$, for the critical bosonic string),
and $\lambda = 1/\epsilon = \sqrt{\alpha'}$ is the string length scale~\cite{Polchinski:1998rq}.
The mutually compatible dynamical fields $\omega_{AB}(\X),\eta_{AB}(\X)$, and $H_{\!AB}(\X)$ are respectively:
the antisymmetric symplectic structure $\omega_{AB}$,
the symmetric polarization (doubly orthogonal) metric $\eta_{AB}$, and
the doubled symmetric metric $H_{\!AB}$, which together define a (dynamical) Born geometry~\cite{Freidel:2013zga,Freidel:2018tkj}.

Quantization renders the doubled ``phase-space'' operators 
$\hat{\X}^A = (\hx^a/\lambda, \htx_b/\lambda)$ inherently non-commutative~\cite{Freidel:2017wst}:
\be
\big[\, \hat{\X}^A,\, \hat{\X}^B\,\big] \,=\, i \omega^{AB}\;.
\label{e:CnCR}
\ee
In this formulation, all effective fields must be regarded a priori as bi-local $\phi(x, \tx)$~\cite{Freidel:2016pls, Freidel:2018apz, Freidel:2021wpl}, subject 
to~\eqref{e:CnCR}, and therefore inherently non-local (yet covariant) in the conventional $x^a$-spacetime. Such non-commutative field theories~\cite{Grosse:2004yu} generically display a mixing between the UV and IR physics, 
with continuum limits defined via a double-scale renormalization group (RG) and self-dual fixed points~\cite{Grosse:2004yu,Freidel:2017xsi}. 
In the current case, the UV and IR mixing occurs between the observable $x^a$-spacetime and the unobservable
$\tx_a$-spacetime.

The metastring offers a new view on quantum gravity by noting that
the world-sheet can be made modular in our formulation, with the doubling of $\tau$ and $\sigma$,
so that  $\hX (\tau, \sigma)$ can be
in general viewed as an infinite dimensional matrix (the matrix indices coming from the Fourier components of the
doubles of $\tau$ and $\sigma$)~\cite{Gopakumar:1994iq,Berglund:2020qcu}.
Then the corresponding metastring matrix model action should look like (with Weyl ordering)
\be
S \sim \int  d \tau\, d \sigma\ \mathrm{Tr}
\left[\,
 \partial_{\tau}   \hX^A \partial_{\sigma} \hX^B (\omega_{AB} (\hX) + \eta_{AB}(\hX)) 
-\partial_{\sigma} \hX^A \,H_{AB} (\hX)\, \partial_{\sigma} \hX^B 
\,\right] 
\;,
\ee
where the trace is over the infinite matrix indices. 
The matrix entries become the natural partonic degrees of freedom of quantum spacetime.
The non-perturbative formulation of quantum gravity is
obtained by replacing $\partial_\sigma$ in the above worldsheet action with a
commutator involving one extra $\hX^{26}$ : 
\be
\partial_{\sigma} \hX^A \quad\to\quad \left[\,\hX^{26},\, \hX^A \,\right]\;,\qquad\qquad
A\,=\, 0,\,1,\,\cdots,\,25\;.
\ee
Therefore, as with the relationship between M-theory and type IIA string theory, a fully interactive and non-perturbative formulation of metastring theory is given 
in terms of a matrix model form of the above metastring action (with $a,b,c=0,1,2,\cdots,25, 26$)
\be
S\sim \int d \tau\ \mathrm{Tr} 
\left(\, 
\partial_{\tau} \hX^a \left[ \hX^b, \hX^c \right] \eta_{abc} (\hX) 
- H_{ac} \left[ \hX^a, \hX^b \right] \left[ \hX^c, \hX^d \right] H_{bd} (\hX)
\,\right) \;,
\ee
where the first term is of the Chern--Simons form, the second term is of the Yang--Mills form, 
and $\eta_{abc}$ contains both $\omega_{AB}$ and $\eta_{AB}$.
In general, we do not need an overall trace if we think of quantum gravity as a pure quantum theory.
Thus, the following matrix model becomes a pure quantum formulation of quantum gravity (viewed as gravitized quantum theory~\cite{Hubsch:2024agh})
\begin{equation}
\mathbb{S}_{\textit{nc}\text{M}}
\;=\; \frac{1}{4\pi}  
\int_{\tau} %
\left(\,
  \partial_{\tau} \hX^{i}
  \left[ \hX^{j}, \hX^{k} \right] g_{ijk} (\hX) 
- \left[ \hX^{i}, \hX^{j} \right]
  \left[ \hX^{k}, \hX^{\ell} \right] 
  h_{ijk\ell} (\X)
\,\right)\;,
\label{metastringAction}
\end{equation}
with $27$ bosonic $\hX$ matrices.\footnote{In this formulation, supersymmetry and its avatars are not fundamental
but emergent~\cite{Freidel:2017xsi}.} 
Within this formulation, both matter and gravitational sectors emerge from the
dynamics of the partonic quanta of quantum spacetime.

In particular, in~\cite{Berglund:2019ctg} 
it has been argued that the generalized geometric formulation of string theory discussed above,~\eqref{metastringAction},
provides an effective description of dark energy, and a de Sitter spacetime.
This is due to the theory's chirality and non-commutatively, as in~\eqref{e:CnCR}, 
doubled realization of the target space, 
and the stringy effective action on the doubled non-commutative spacetime $(x^a,\tx_a)$,
which leads to the effective action
\be
S_{\text{eff}}^{\textit{nc}}
\;=\;
\int_x \int_{\tilde{x}} \text{Tr} \sqrt{g(x,\tx)}\, 
\Bigl[\,
R(x,\tx) + L_m(x,\tx) + \cdots
\,\Bigr]\;,
\label{e:ncEH}
\ee
where the ellipses denote higher-order curvature terms induced by string theory, and
$L_m$ is the matter (both visible and dark) Lagrangian put in by hand.
This result can
be understood as a generalization of the famous calculation in string theory~\cite{Polchinski:1998rq}.  
Owing to~\eqref{e:CnCR}, we have
\be
\bigl[\,\hx^a, \,\htx_b\,\bigr] \,=\, 2\pi i\,\lambda^2\,\delta^a_b\;,\qquad 
\bigl[\,\hx^a, \,\hx^b \,\bigr] \,=\, 
\bigl[\,\htx_a,\,\htx_b\,\bigr] \,=\, 0\;,
\ee
where $\lambda$ denotes the fundamental length scale, such as the Planck scale,
and $\epsilon = 1/\lambda$ is the corresponding fundamental energy scale, while the string tension is 
$\alpha' = \lambda/\epsilon = \lambda^2$.
Thus $S_{\text{eff}}^{nc}$ expands into numerous terms with different powers of $\lambda$, 
which upon $\tx$-integration, and from the $x$-space vantage point, produce various effective terms.
To lowest (zeroth) order of the expansion in the non-commutative parameter $\lambda$
of $S_{\text{eff}}^{\textit{nc}}$ takes the form:
\bea
S_{d=4} & = & - \int_{x}\int_{\tx} \sqrt{-g(x)} \sqrt{-\tilde{g}(\tx)} 
\,\Bigl[\, 
  R(x) + \tilde{R}(\tx)
\,\Bigr]
\cr
& = & -\int_{x}\sqrt{-g(x)}\left[
R(x)\int_{\tx}\sqrt{-\tilde{g}(\tx)} 
+\int_{\tx}\sqrt{-\tilde{g}(\tx)}\;\tilde{R}(\tx)
\right]
\;,
\label{e:TsSd}
\eea
a result which first was obtained almost three decades ago, effectively neglecting $\omega_{AB}$ 
by assuming that $[\,\hat x^a,\,\htx_b\,] = 0$~\cite{Tseytlin:1990hn}. 
In this leading limit, the $\tx$-integration in the first term of~\eqref{e:TsSd} defines the gravitational constant $G_N$, 
\be
1/G_N \;\sim\; \int_{\tx} \sqrt{-\tilde{g}(\tx)} \;,
\ee
and in the second term produces a {\it positive} cosmological constant $\Lambda >0$ (dark energy)
\be
\Lambda/G_N \;\sim\; \int_{\tx} \sqrt{-\tilde{g}(\tx)}\,\tilde{R}(\tx) \;.
\label{LambdaOverGN}
\ee
Thus the weakness of gravity is determined by the size of the canonically conjugate dual $\tx$-space, 
while the smallness of the cosmological constant is given by its curvature $\tilde{R}$. 
Ref.~\cite{Berglund:2019ctg} also discusses a see-saw formula for the cosmological constant, as well
as its radiative stability in the underlying general framework of a
non-commutative generalized geometric phase-space formulation of string 
theory~\cite{Freidel:2013zga, Freidel:2015uug, Freidel:2016pls, Freidel:2017xsi, Freidel:2017wst}, 
which is non-local but covariant.

To summarize, a non-perturbative formulation of
quantum gravity can be given in terms of a doubled matrix model,~\eqref{metastringAction}, 
in which everything is built out of partonic degrees of freedom represented by the entries of doubled matrices $\hX$. 
In the leading term of the effective spacetime description, 
dark energy is realized as a dynamical geometry of the dual spacetime, and consequently, 
is inherently time-dependent.

\section{Infinite statistics}\label{sec:infinitestatistics}

Here we summarize the essential facts about infinite statistics (quantum distinguishable, or quantum Boltzmann statistics). The $q$-deforemed algebra of harmonic oscillators can be written
\be
\big[\,\hat{a}_i^{\phantom{\dagger}},\, \hat{a}_j^\dagger\,\big]_q 
\;=\; \hat{a}_i^{\phantom{\dagger}} \hat{a}_j^\dagger - q\,\hat{a}_j^\dagger \hat{a}_i^{\phantom{\dagger}} 
\;=\; \delta_{ij} \;. 
\label{eq:qdef}
\ee
Setting $q=1$ is the bosonic case, and~\eqref{eq:qdef} describes the usual quantum harmonic oscillator.
Similarly, setting $q=-1$ in~\eqref{eq:qdef} is the fermionic case in which the commutator is replaced by an anti-commutator.
These yield, respectively, particles with Bose--Einstein and Fermi--Dirac statistics~\cite{Greenberg:1989ty}.

Setting $q=0$ in~\eqref{eq:qdef} also yields a familiar statistics, \textit{viz.}, infinite statistics, corresponding to the Cuntz oscillators~\cite{voiculescu1992free, Gopakumar:1994iq}.
Explicitly, Cuntz oscillators satisfy the algebra
\be
\hat{a}_i^{\phantom{\dagger}} \ket{0} \;=\; 0 \;, \qquad 
\hat{a}_i^{\phantom{\dagger}} \hat{a}_j^\dagger \;=\; \delta_{ij} \;, \qquad 
\sum_i \hat{a}_i^\dagger \hat{a}_i^{\phantom{\dagger}} \;=\; \mathbf{1} - \ket{0}\bra{0}\;.
\ee
Because there are no further relations, the raising and lowering operators neither commute nor anti-commute.
This means the ordering is important: $\hat{a}_i^\dagger \hat{a}_j^\dagger \ne \pm\, \hat{a}_j^\dagger \hat{a}_i^\dagger$.
Respecting this property, the number operator is given by a sum on words~\cite{Greenberg:1989ty,Halpern:2001hg}:
\be
\hat{N} \;=\; \sum_{k=1}^\infty \sum_{i_1} 
\hat{a}_{i_1}^\dagger \Big( \sum_{i_2} \hat{a}_{i_2}^\dagger \ldots \Big( \sum_{i_k} \hat{a}_{i_k}^\dagger \hat{a}_{i_k}^{\phantom{\dagger}} \Big) \ldots \hat{a}_{i_2}^{\phantom{\dagger}} \Big) \hat{a}_{i_1}^{\phantom{\dagger}} \;. 
\label{eq:numop}
\ee
Even in the single oscillator case, we have
\be
\hat{N}_1 
\;=\; \sum_{k=1}^\infty (\hat{a}^\dagger)^k (\hat{a})^k 
\;=\; \frac{\hat{a}^\dagger \hat{a}}{1-\hat{a}^\dagger \hat{a}} 
\;.
\ee
The summand $(\hat{a}^\dagger)^k (\hat{a})^k$ counts the presence of a particle in any given excitation.
When there are more oscillators, the number operator~\eqref{eq:numop} accounts for distinguishable particles, which then satisfy Boltzmann statistics.

The Cuntz algebra is intrinsic to matrix theories.
If we have $n$ independent random matrices, working in the large-$N$ limit, the Fock space obtained from acting the matrices on the vacuum, \textit{i.e.}, by taking $\hat{M}_1\hat{M}_2\ldots |0\rangle$, is isomorphic to the Cuntz algebra~\cite{Gopakumar:1994iq}.
This operation may, for instance, be natural in the Matrix model for M-theory~\cite{Banks:1996vh}.
In the BFSS Matrix model, gravitons are bound states of D$0$-branes.
The gravitational interaction and the geometry of spacetime arise from off-diagonal matrix elements corresponding to open string degrees of freedom stretching between branes.
The D$0$-branes (the partonic degrees of freedom) obey infinite statistics.

The quantum statistics associated to the Cuntz algebra can also be realized in other macroscopic settings in quantum gravity.
In~\cite{Strominger:1993si}, the statistics of extremal black holes is described in terms of infinite statistics.
In~\cite{Jejjala:2007hh, Jejjala:2007rn}, the authors argue that dark energy quanta arising from the Matrix theory framework obey infinite statistics and follow the Wien distribution.
This latter observation is an impetus for the present work.

\bibliographystyle{JHEP}
\bibliography{refs}

\end{document}